\documentclass[letter,8pt]{article}
\usepackage{graphicx}
\usepackage{epsfig}
\usepackage{subfigure}
\usepackage{braket}
\usepackage{pslatex,latexsym}
\usepackage{bm,float,lipsum}
\usepackage{amsmath}
\usepackage{amssymb}
\usepackage{amsfonts}
\usepackage{mathrsfs}
\usepackage{color}
\usepackage{rotating}
\usepackage{array}
\usepackage{amsfonts}
\usepackage{booktabs}
\usepackage{threeparttable}
\usepackage{multirow}
\usepackage{times}
\usepackage{longtable}
\usepackage{chngpage}
\usepackage{tabulary}
\usepackage[hmargin=2.5cm,vmargin=3cm]{geometry}
\usepackage{indentfirst}
\usepackage[labelfont=bf]{caption}
\usepackage{ulem}

\DeclareCaptionLabelSeparator{vline}{. }
\usepackage{authblk}
\begin{document}
\title{Quantitative assessment of the effects of resource optimization and ICU admission policy on COVID-19 mortalities}
\author[a]{Ying-Qi Zeng \thanks{These authors contributed equally to this work.}}
\author[b]{Lang Zeng \thanks{These authors contributed equally to this work.}}
\author[b,c]{\\Ming Tang \thanks{Corresponding author: tangminghan007@gmail.com}}
\author[d]{Ying Liu}
\author[b]{Zong-Hua Liu}
\author[e]{Ying-Cheng Lai \thanks{Corresponding author: Ying-Cheng.Lai@asu.edu}}

\affil[a]{School of Communication and Electronic Engineering, East China Normal University, Shanghai 200241, China}
\affil[b]{State Key Laboratory of Precision Spectroscopy and School of Physics and Electronic Science, East China Normal University, Shanghai 200241, China}
\affil[c]{Shanghai Key Laboratory of Multidimensional Information Processing, East China Normal University, Shanghai 200241, China}
\affil[d]{School of Computer Science, Southwest Petroleum University, Chengdu 610500, P. R. China}
\affil[e]{School of Electrical, Computer and Energy Engineering, Arizona State University, Tempe, AZ 85287, USA}

%\author{Ying-Qi Zeng}
%\thanks{These authors contributed equally to this work.}
%\affiliation{School of Communication and Electronic Engineering, East China Normal University, Shanghai 200241, China}
%
%\author{Lang Zeng}
%\thanks{These authors contributed equally to this work.}
%\affiliation{State Key Laboratory of Precision Spectroscopy and School of Physics and Electronic Science, East China Normal University, Shanghai 200241, China}
%
%\author{Ming Tang} \email{tangminghan007@gmail.com}
%\affiliation{State Key Laboratory of Precision Spectroscopy and School of Physics and Electronic Science, East China Normal University, Shanghai 200241, China}
%\affiliation{Shanghai Key Laboratory of Multidimensional Information Processing, East China Normal University, Shanghai 200241, China}
%
%\author{Ying Liu}
%\affiliation{School of Computer Science, Southwest Petroleum University, Chengdu 610500, P. R. China}
%
%\author{Zonghua Liu}
%\affiliation{State Key Laboratory of Precision Spectroscopy and School of Physics and Electronic Science, East China Normal University, Shanghai 200241, China}
%
%\author{Ying-Cheng Lai} \email{Ying-Cheng.Lai@asu.edu}
%\affiliation{School of Electrical, Computer and Energy Engineering, Arizona State University, Tempe, AZ 85287, USA}

\date{\today}
\maketitle
\begin{abstract}
It is evident that increasing the intensive-care-unit (ICU) capacity and giving priority to admitting and treating younger patients will reduce the number of COVID-19 deaths, but quantitative assessment of these measures has remained inadequate. We develop a comprehensive, non-Markovian state transition model, which is validated through accurate prediction of the daily death toll for two epicenters: Wuhan, China and Lombardy, Italy. The model enables prediction of COVID-19 deaths in various scenarios. For example, if treatment priorities had been given to younger patients, the death toll in Wuhan and Lombardy would have been reduced by 10.4\% and 6.7\%, respectively. The strategy depends on the epidemic scale and is more effective in countries with a younger population structure. Analyses of data from China, South Korea, Italy, and Spain suggest that countries with less per capita ICU medical resources should implement this strategy in the early stage of the pandemic to reduce mortalities.
\end{abstract}
\section*{Introduction}

The continuous and recently accelerated spreading of COVID-19 in more than 145
different countries and regions of the world has placed an unprecedented
burden on the corresponding healthcare systems. As of December 1, 2020, there
have been more than 63.3 million confirmed cases with over 1.4 million deaths
reported, and the daily number of deaths~\cite{label1} has exceeded 8000. To
mobilize the available medical resources to the maximum degree to reduce the
COVID-19 fatalities has become the top priority in many hospitals and
healthcare facilities. In a hospital, there are two types of care resources for
inpatients: General Ward (GW) and Intensive Care Unit (ICU) beds, with the
latter currently serving mostly COVID-19 patients. Increasing the ICU beds
would undoubtedly reduce the COVID-19 fatalities. When a hospital is
overwhelmed with COVID-19 patients so that the ICU beds are in a serious
shortage, a selective ICU admission policy must be invoked to allow specific
groups of patients to receive the ICU care. In this regard, from a sheer
mathematical point of view, giving priority to the young age group will reduce
the fatality rate, as patients in this group have a better chance of being
cured. Stretching the ICU resources to their limit and implementing selective
ICU admission policy are becoming inevitable and even absolutely necessary in
many parts of the world as the COVID-19 cases have continued to skyrocket in
recent months.

It is intuitively evident that enhancing ICU capabilities and selectively
admitting patients into ICU can reduce the COVID-19 mortalities, but we lack
a modeling framework to quantitatively assess and characterize their effects.
The main goal of this paper is to address this issue that is critically
important to maximizing the usage and effectiveness of the limited medical
resources to minimize COVID-19 deaths.

There have been intense modeling efforts on early warning, prevention and
control of the COVID-19 pandemic~\cite{label2,label3,label4,label5,label6,label7,label8,label9,label36,label37}.
For example, Kraemer et al.~\cite{label2}
found that travel restriction in the early stage of the COVID-19 outbreak can
effectively prevent the infection imported from known sources. Once
cases begin to spread in the community, the contribution of newly imported
cases tends to diminish, requiring a set of control measures including travel
restrictions, detection, tracking, and isolation to mitigate the pandemic.
Kissler et al.~\cite{label3} established a SARS-CoV-2 transmission model with
seasonal variations, immune duration, and cross immunization, where the peak
of SARS-CoV-2 infection is assumed to be low in spring and summer and a larger
epidemic breakout can occur in autumn and winter. A finding was that, with
respect to imposing one time social distancing, intermittent social alienation
measures can prevent the overload of public health resources. Hao
et al.~\cite{label4} found that the new coronavirus has two characteristics:
high infectivity and concealment. If 87\% of the infected individuals
are not detected, without any prevention and control measures, after 14
consecutive days of zero confirmed cases, the probability of a second wave
of epidemics will be 32\%. If only 53\% of the infected are undetected, the
probability of an epidemic rebound will drop to 6\%. Premature removal of
prevention and control measures will greatly increase the possibility of a
second outbreak of the epidemic. Long et al.~\cite{label5,label6,label7}
developed a time-delay transmission model that can accurately simulate and
predict the epidemic development in countries or regions, and the model
enables evaluation of the impact of virus detection and social distancing
intensity on the epidemic, as well as an accurate estimate of the time zero
point of the epidemic in various countries and regions. In particular,
it was found~\cite{label7} that community-based transmission occurred in
Europe and in the United States in early January 2020. The study of Flaxman
et al.~\cite{label8} revealed that the outbreak interventions implemented in
11 European countries have reduced about 1.3 million deaths and these
measures are enough to bring down the basic reproduction number $R_0$ to
less than one. Dehning et al. ~\cite{label9} studied the spread of COVID-19
in Germany by combining epidemiological modeling with Bayesian inference, and
found that the change point of the effective growth rate of new infection is
closely related to the time point when the intervention measures were imposed.

There have also been recent studies on the interplay between the medical
resources and COVID-19 mortality. Particularly, in comparison with the common
influenza virus, a higher proportion of the patients infected with SARS-CoV-2
need to be hospitalized. The aggravated growth in the number of infected
individuals in recent months has stretched the capacities of the medical
systems in many countries and regions to their limit. A shortage of medical
resources, such as the respiratory support devices, will result in a large
number of severely ill patients not getting timely and effective treatment,
exposing them to a greater risk of mortality. Ferguson
et al.~\cite{label10,label11} evaluated the demand for medical resources of
the non-pharmaceutical interventions type, predicting that, even when the most
effective mitigation measures are implemented, the numbers of hospital GW and
ICU beds in the UK still need to be expanded by more than eight times to meet
the needs of patient care. Miller et al.~\cite{label12} used the age-specific
mortality and demographic data to predict the cumulative COVID-19 cases and
medical resource burden in different regions of the United States, pointing
out that medical resources are relatively scarce in remote areas. When the
medical system is overloaded, the nursing standard for the patients will be
compromised, which would affect the treatment outcome. Without proper and
rigorous care, the condition of critically ill patients with COVID-19 will
further deteriorate, and the shortage of ICU facilities will aggravate the
high mortality rate of such population~\cite{label13,label14}. These
studies~\cite{label10,label11,label12,label13,label14} made evident the
potential significant impact of the medical resource availability on the
COVID-19 mortality.

In this paper, we develop a comprehensive state transition model to predict
the COVID-19 death and its evolution over time for a regional healthcare
system with limited medical resources and selective ICU admission policy. The
typical systems are a city or a region, e.g., Wuhan city in China or the
Lombard region of Italy. Realistic time delays associated with various state
transitions are fully incorporated into the model, rendering it non-Markovian
to better describe the real world situations. The model is based on and
validated with empirical data such as the number of confirmed cases, clinical
data, demographics, and the amount of medical resources, and it enables such
questions to be addressed as: if the ICU resources were deployed certain
days earlier or if the ICU capacity was doubled or tripled, how much reduction
in the death toll would be achieved? The findings of this paper can be best
described with concrete numbers. For example, for Wuhan city, if the ICU
resources had been deployed a week in advance or if the number of ICU beds
had been doubled, the death toll would have been reduced by 5\% or 13\%,
respectively. For Lombardy, the corresponding numbers are similar:
3\% or 14\%. For both Wuhan and Lombardy, tripling the ICU capacity would
have resulted in a 21\% reduction in the mortalities. With respect to
selective ICU admission policy, prioritizing younger patients would have
reduced the death toll in Wuhan and Lombardy by 10.4\% and 6.7\%, respectively.
As illustrated by the exemplary numbers, our model provides a framework to
predict COVID-19 deaths in arbitrary scenarios. Because the model has been
fully validated with real data, we expect the model predictions to be reliable
and accurate. A further analysis of the data from China, South Korea, Italy
and Spain indicates that the age-selective admission strategy is more effective
in countries with a younger age structure, while the countries with less per
capita ICU medical resources should implement the selective admission strategy
in the early stage of the epidemic to suppress mortalities.

\section*{Model}

\begin{figure} [ht!]
\center
\includegraphics[width=0.8\linewidth]{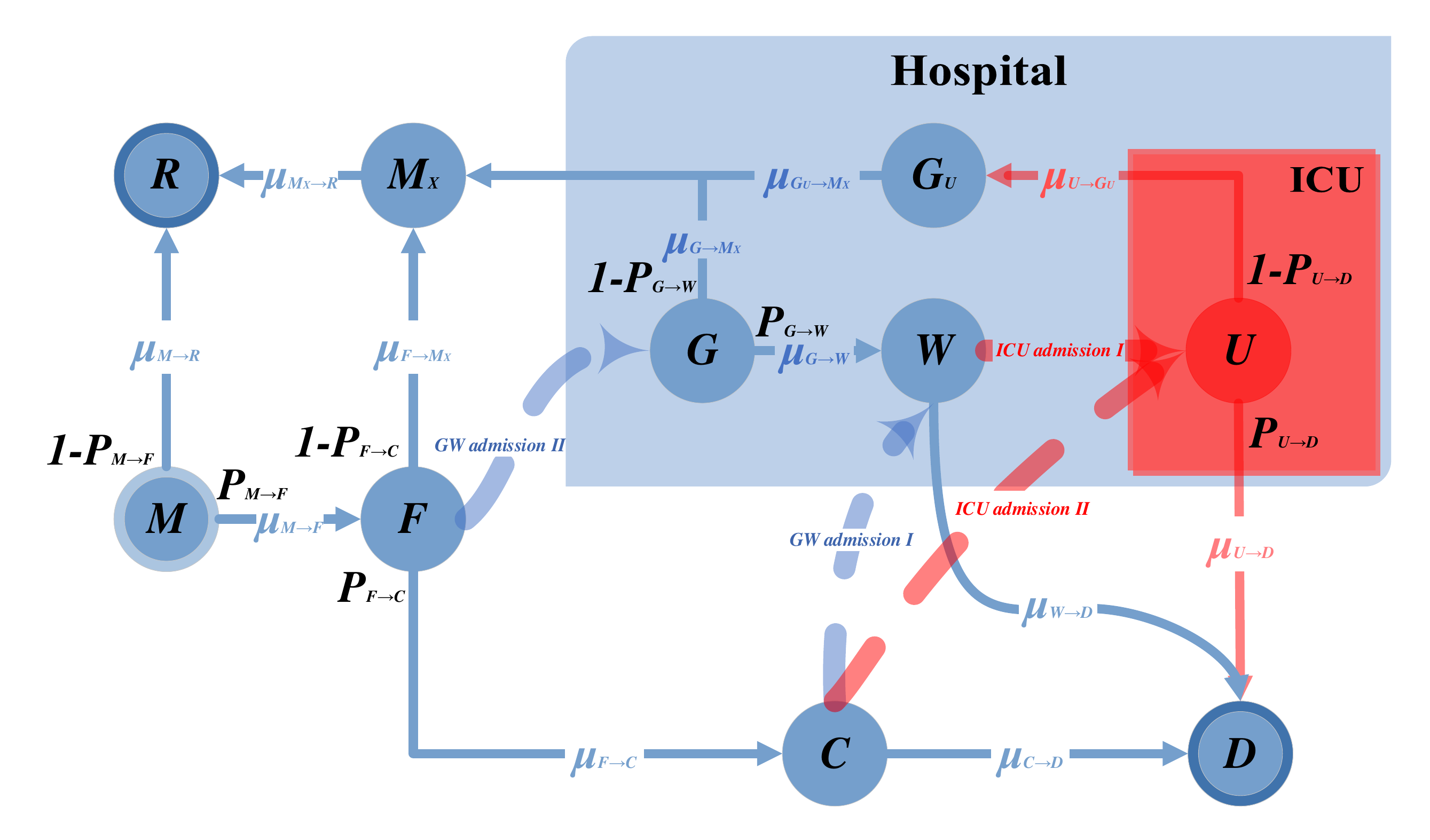}
\caption{ {\em Schematic diagram of the proposed COVID-19 patient admission
model under limited medical resources.} Each node represents a possible
state of the patients. The specific definition of each state and the
transition paths among them (light blue and light red arrows with time
representing the average delay of the transition) are described in detail in
the text. The various quantities $P_{\cdot\rightarrow\cdot}$ denote the
fractions of the state transition paths, and the time delays follow
a normal distribution. The light red dotted arrows indicate the two admission
paths of ICU, where ICU admission I and II have the first and second priority,
respectively. The light blue dotted arrows indicate the two treatment paths of
GW, which have a lower priority than ICU admission: GW admission I and II have
the third and fourth priority, respectively.}
\label{fig:SD}
\end{figure}

Our COVID-19 patient admission model under limited medical resources
has ten dynamical states, as shown in Fig.~\ref{fig:SD}. It describes the
basic transition process of a diagnosed individual's state from onset to
recovery or death. Time delays associated with the various state transitions
are taken into account, rendering non-Markovian the dynamical evolution. In
particular, state M denotes that a patient is infected with COVID-19, state
F means that the patient needs to be hospitalized but is currently under home
isolation or centralized isolation, state C describes that the patient needs
ICU support treatment but is currently isolated, states G and U indicate an
inpatient occupying a GW and an ICU bed, respectively, a patient in state W
currently occupies a GW bed but needs ICU treatment, a patient in the $G_U$
state is transferred out of ICU who does not need to occupy a GW bed, a
patient in the $M_x$ state is discharged from the hospital with lessened
symptoms, and the R and D states represent two clinical outcomes: recovery
and death.

The medical resources can be divided into two categories: GW and ICU resources
denoted as $Q_G$ and $Q_U$, respectively, which can be measured, e.g., by the
corresponding numbers of beds. That the resources are limited is
modeled, as follows. when the GW beds are all used up, new patients will no
longer be admitted. When all ICU beds have been occupied, patients who require
ICU care either continue to wait at GW (W-state) or continue to wait
outside the hospital (C-state). A common hospital admission policy is
``first-come first-serve'' (FCFS), which gives priority to patients
who have had a long waiting time. The admission policies of GW and ICU thus
consist of four admission paths with different priorities: GW admission I/II
and ICU admission I/II, where ICU admission I for W-state patients has the
highest priority, followed by ICU admission II for C-state patients, GW
admission I for the remaining C-state patients in GW, and GW admission II for
the F-state patients. The priorities define the execution order of the
corresponding paths. For M-state patients, two different transitions can
occur: some patients deteriorating into the F-state so as to require
hospitalization and others slowly recovering into the R-state. For patients
in the F-state, there are three types of transitions: (i) when the GW
resources are available, the patients are admitted to the hospital through GW
admission II, (ii) a proportion of the remaining F-state patients deteriorate
to C-state, requiring ICU treatment, and (iii) the rest are cured so they
switch to the $M_X$-state. Patients in state C enter ICU through ICU admission
II to transition into U-state, or through GW admission I to enter GW to become
W-state, while the rest are in D-state. For G-state patients, some switch to
$M_X$-state and no longer occupy GW resources, while others deteriorate to
the W state, requiring ICU care. W-state patients will enter ICU through ICU
admission I, and the remaining will be in the D state. Some of the U-state
patients occupying ICU resources will transition to the  $G_U$ state, and the
others will switch to the D state. For patients in the $G_U$ state, the
symptoms are relieved after a period time and they enter the $M_X$ state.
Patients in the $M_X$ state recover to the R state after a period of time.

Let $\bigtriangleup Q_G$ and $\bigtriangleup Q_U$ denote the changes in the
GW and ICU resources, respectively. A decrease in the GW resources is the sum
of the number of patients admitted through the GW admission I/II pathway,
and an increase is the sum of the number of ICU admission I, the number of
patients in G (W)-state cured (or died), and the amount of newly deployed GW
resources. Likewise, a decrease in ICU resources is the sum of the number of
patients admitted through ICU admission I/II pathway, and an increase is the
sum of the number of patients cured (or died) in state U and any newly
deployed ICU resources. The state dynamical evolution described by a set of
difference equations with time delays characterizing the state ages
[Supplementary Note 1 (SN 1)].

\section*{Results}

We validate our ten-state model by simulating the trend of the death tolls
in Wuhan and Lombardy, using the daily number of confirmed cases, clinical
and demographic data. The model then enables us to assess the overall effects
of varying the timing of resource deployment and its amount on the COVID-19
mortalities. At a more detailed level, we divide the patients into several
age groups and calculate the death-toll trend in each age group. This
allows us to assess the impact of resource allocation scheme of different age
groups on the number of deaths, and to obtain the optimal ICU admission
strategy in terms of the age structure and outbreak scale with limited
medical resources.

\subsection*{Impact of limited medical resource deployment}

\subsubsection*{Model validation}

To simulate the trend of the death tolls in Wuhan and Lombardy using the
the second-order difference equations in our modeling framework, three types
of data are required: time delays associated with state transitions, patient
morbidity data, and local medical resource deployment data. The details of
these data are described in SN2 and Supplementary tables S1-S5. which include
information such as the average number of days in the transition delay from
M-state to F-state and the average mortality rate of patients in ICU. Most of
the data were obtained through references and official
reports~\cite{label15,label16,label17}. As detailed in SN2, the
values of some model parameters need to be estimated in an optimal way, e.g.,
at the 95\% confidence level, through model simulation and empirical data such
as the average fraction of the patients switching from G-state to W-state,
denoted as $|P|_{G\rightarrow W}$, and that from F-state to C-state, denoted
as $|P|_{F\rightarrow C}$. We use the cumulative days of insufficient GW and
ICU resources to quantify the level of stress on the healthcare system,
defined as the GW overload days $O_G$ and ICU overload days $O_U$
(see SN3 -  medical system stress level indicators).

\begin{figure} [ht!]
\includegraphics[width=0.8\linewidth]{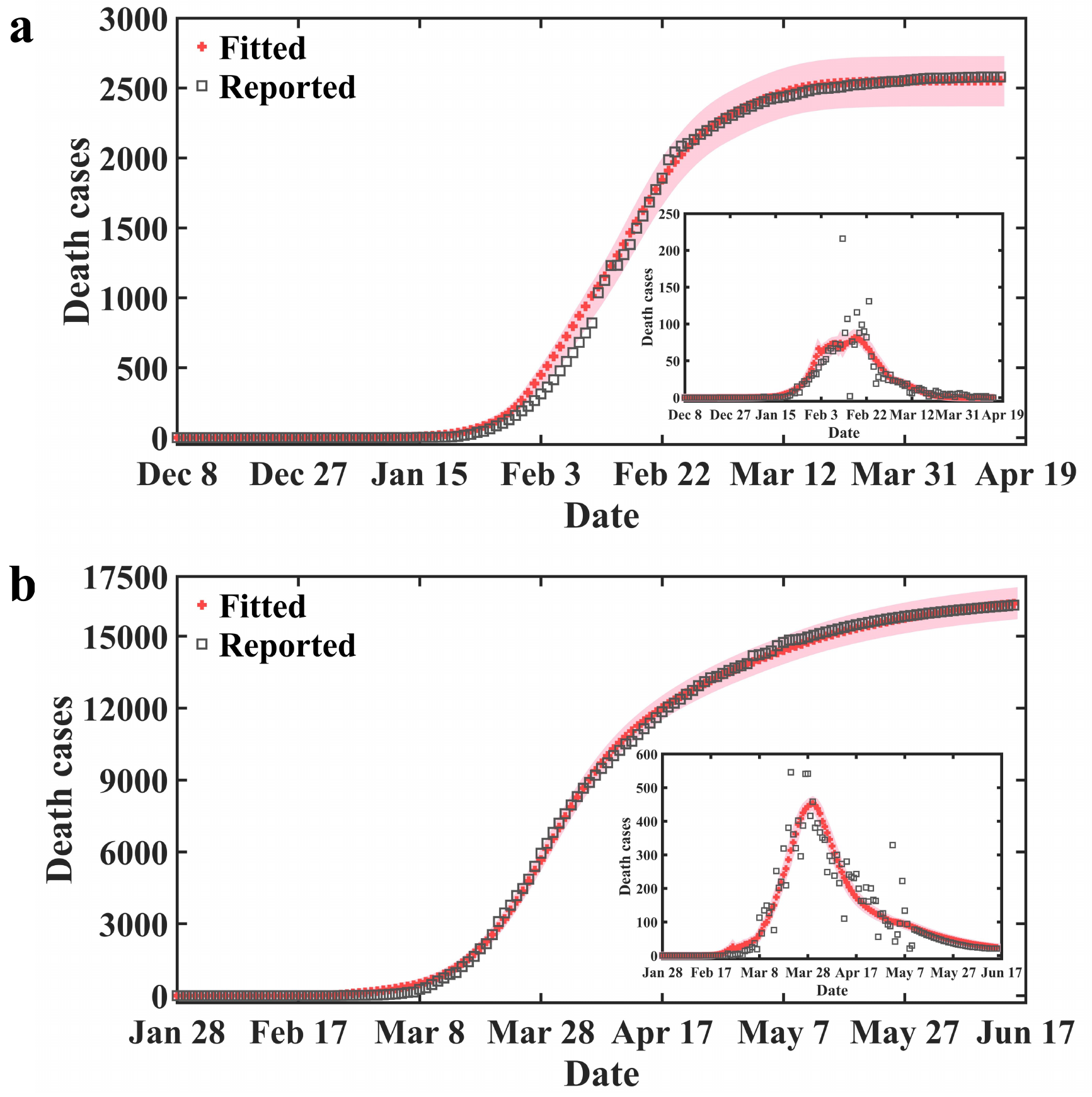}
\caption{ {\em Simulation results of the daily evolution of death tolls
for the two COVID-19 epicenters.} (a,b) Actual data of cumulative death toll
with time in Wuhan and Lombardy, respectively. The insets show the daily
numbers of new deaths. The open black squares and red crosses represent the
actual data and simulation results, respectively. The pink shaded area
marks the 95\% confidence interval. There is a good agreement between the
simulated daily number of deaths and the actual data, validating our ten-state
model.}
\label{fig:SM}
\end{figure}

With the optimal parameter values and empirical data as inputs to the model,
we simulate the cumulative death toll and new deaths over time in Wuhan
(from December 8, 2019 to April 15, 2020) and Lombardy (from January
28 to June 14, 2020), as shown in Fig.~\ref{fig:SM}. It can be seen that the
model predicted trends of the cumulative and daily new death tolls agree with
the actual data very well, validating the model. The model also gives that
the values of $O_G$ are 14 and 17 days for Wuhan and Lombardy, respectively,
and the corresponding values of $O_U$ are 49 and 74 days.

\subsubsection*{Deployment of medical resources}

Since the onset of COVID-19 pandemic, many countries (most notably the USA) and
regions have missed the best time window to control the spreading. At the
present, there is a skyrocketing increase in the demand for ICU beds and
medical resources such as ventilators and other special medical devices.
Many hospitals and healthcare facilities have reached or will soon reach the
limit of their operating capacities.

Optimizing the usage of medical resources by deploying them as early as
possible and augmenting them as much as possible are key to reducing the
mortality rate. Let DT and RI denote the two key parameters: the deployment
time and the resource input, where $\mbox{DT}=0$ and $\mbox{RI}=1$ represent
the actual deployment time and the available normalized amount of resources,
respectively. If the resources are deployed seven days ahead of the actual
time, we have $\mbox{DT}=-7$. Likewise, if the resources are doubled (e.g.,
twice as many ICU beds as in the actual case), we have $\mbox{RI}=2$ (see
SN3 for COVID-19 special medical resource deployment plans). A virtue
of our modeling framework lies in its ability to provide a quantitative
picture of the dependence of the mortalities on the two key parameters.

To uncover the impact of deploying medical resources on the patient mortality
rate in a concrete way, we consider the following three scenarios: (i) varying
the GW resources only, (ii) varying ICU resources only, and (iii) varying
both resources simultaneously. Some representative results are presented below
(more results in SN4).

\begin{figure} [ht!]
\includegraphics[width=0.9\linewidth]{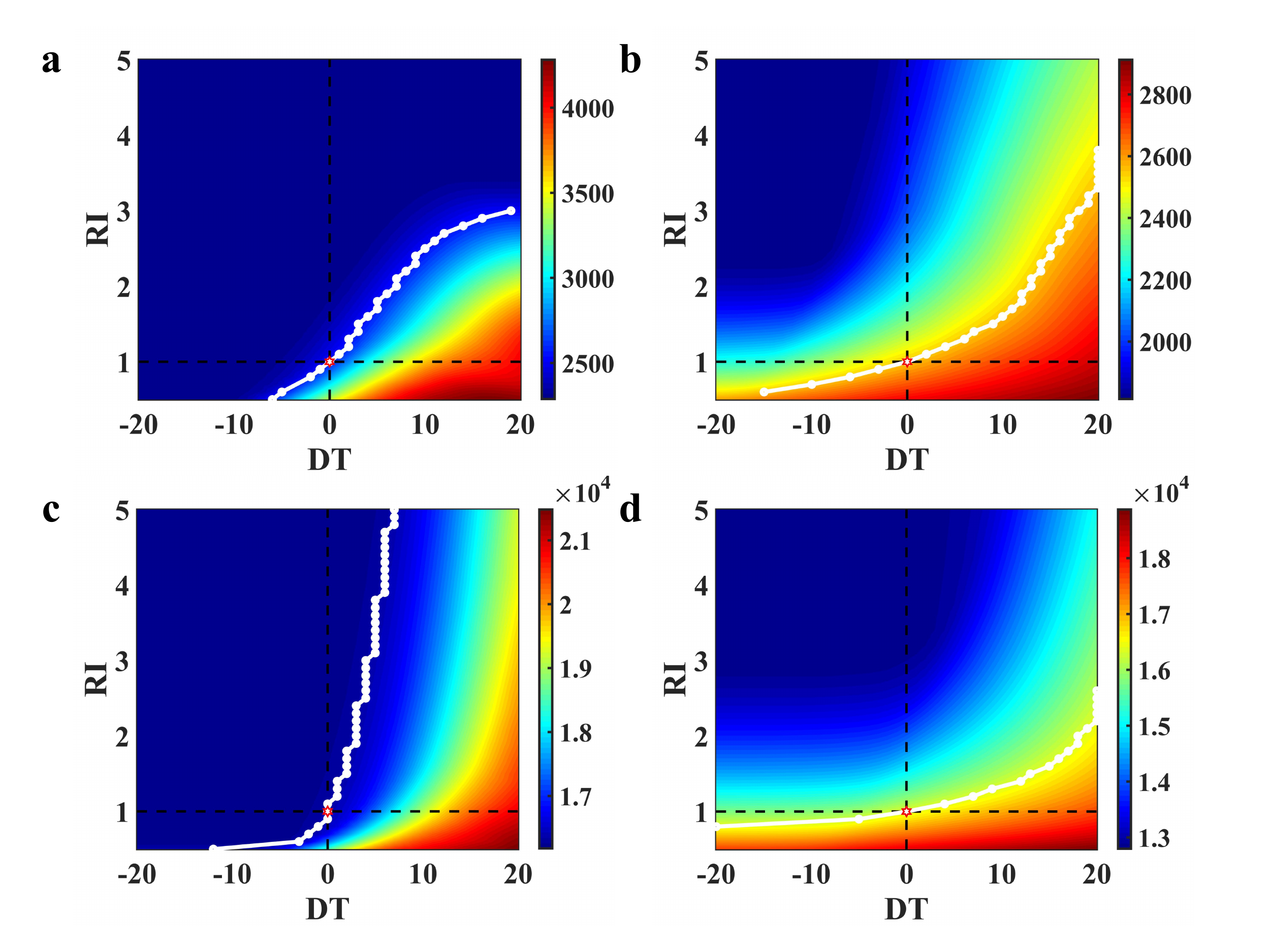}
\caption{ {\em Impact of timing of medical resource deployment and resource
input on COVID-19 mortalities.} Each panel shows the number of deaths (color
coded) in the parameter plane of deployment time and resource input: panels
(a,b) are for Wuhan, China, and (c,d) are for Lombardy region, Italy. In
panels (a,c), the ICU resource amount is fixed and the resource that varies is
GW. In panels (b,d), the GW resource amount is fixed and the resources that
vary are ICU. The red hexagon indicates the true death toll under the actual
resource deployment, and the white circles mark the contour of the actual
death toll in the parameter plane.}
\label{fig:PD}
\end{figure}

Figure~\ref{fig:PD}(a) shows, for Wuhan, the impact of varying GW resource
deployment on the number of deaths for fixed ICU resource deployment, where the
red hexagon indicates the true death toll under the actual resource deployment.
%If the original deployment is delayed by one week or the resources investment is reduced by half, the death toll will increase by 32\% or 33\%, respectively.
If the GW resources had been deployed one week in advance or if its amount had
been doubled, the death toll would have been reduced by 10\%. While remarkable,
this 10\% figure is about the maximum reduction that can be achieved by varying
the GW resource deployment. For example, deploying the GW resources earlier
than one week or an increase in its investment over three times would not
reduce the death toll further. The maximum reduction in the deaths that can
be achieved is determined approximately by the contour of the actual death toll
- the curve connecting the white circles Fig.~\ref{fig:PD}(a). These results
indicate that deploying the GW resources earlier and/or augmenting them have
only limited effects on reducing the deaths.

In contrast, deploying the ICU resources earlier and/or augmenting them can
be more effective at reducing the deaths in Wuhan, as shown in
Fig.~\ref{fig:PD}(b), where the GW resource deployment is fixed.
%It can be seen that, if the deployment had been delayed by one week or if its amount had been reduced by half, the death toll will increase by 4\% and 9\%, respectively.
If the ICU resources had been deployed one week ahead of the actual time or if
the ICU resource amount had been doubled, the death toll would have been
decreased by 5\% and 13\% respectively. While these figures are similar to
those that would have been achieved with the same adjustment in the GW
resources [Fig.~\ref{fig:PD}(a)], deploying the ICU beds earlier or increasing
their number can have a much more significant effect on the number of deaths.
For example, if the number of ICU beds had been tripled, the death toll would
have been reduced by 21\%! The maximum reduction in the deaths is achieved for
$\mbox{DT}=-14$ and $\mbox{RI}=2.5$, which is about 30\% (about three times
more effective than that which can be achieved by stretching the GW resources
and their deployment). The corresponding ICU overload days would have been
decreased to less than seven days (comparing with the actual 49 days for Wuhan).
Similar results are obtained for Lombardy. We find that adjusting the GW
resource deployment would have only limited effects in reducing the death
toll. For example, as shown in Fig.~\ref{fig:PD}(c), if the GW resources had
been deployed one week or two weeks earlier or if the number of GW beds had
been doubled or tripled, the reduction in the deaths would have been only
about 1\%. This indicates that the deployment of the GW resources in Lombardy
was timely and its amount was appropriate, a fact that can also be seen from
the value of the GW overload days $O_G$: to make it less than a week, the
resources would need to be deployed only five days earlier or the number of GW
beds would need to be only 1.4 times higher. In contrast, deploying the ICU
resources earlier or augmenting them would have been much more effective at r
educing the deaths. For example, if the number of ICU beds had been tripled,
the death toll would have been reduced by 21\%, as shown in
Fig.~\ref{fig:PD}(d). The maximum reduction that can be achieved is about
22\% and $O_U$ can be decreased to less than a week.

SN3 presents results on the impact of varying the GW and ICU resource
deployment simultaneously on COVID-19 death toll for both Wuhan and Lombardy.

\subsection*{Admission strategy based on age groups}

To improve the efficiency of treatment and to reduce the mortality rate, it is
essential and imperative to allocate the available resources as reasonably
as possible because as they are not unlimited. For COVID-19, the
hospitalization and mortality rates of elderly patients are higher than those
of younger patients~\cite{label18}. A common strategy is to divide the patients
into distinct age groups. Our modeling framework provides a rigorous way to
calculate the mortalities for different age groups.

\subsubsection*{Age groups}

\begin{figure} [ht!]
\includegraphics[width=0.9\linewidth]{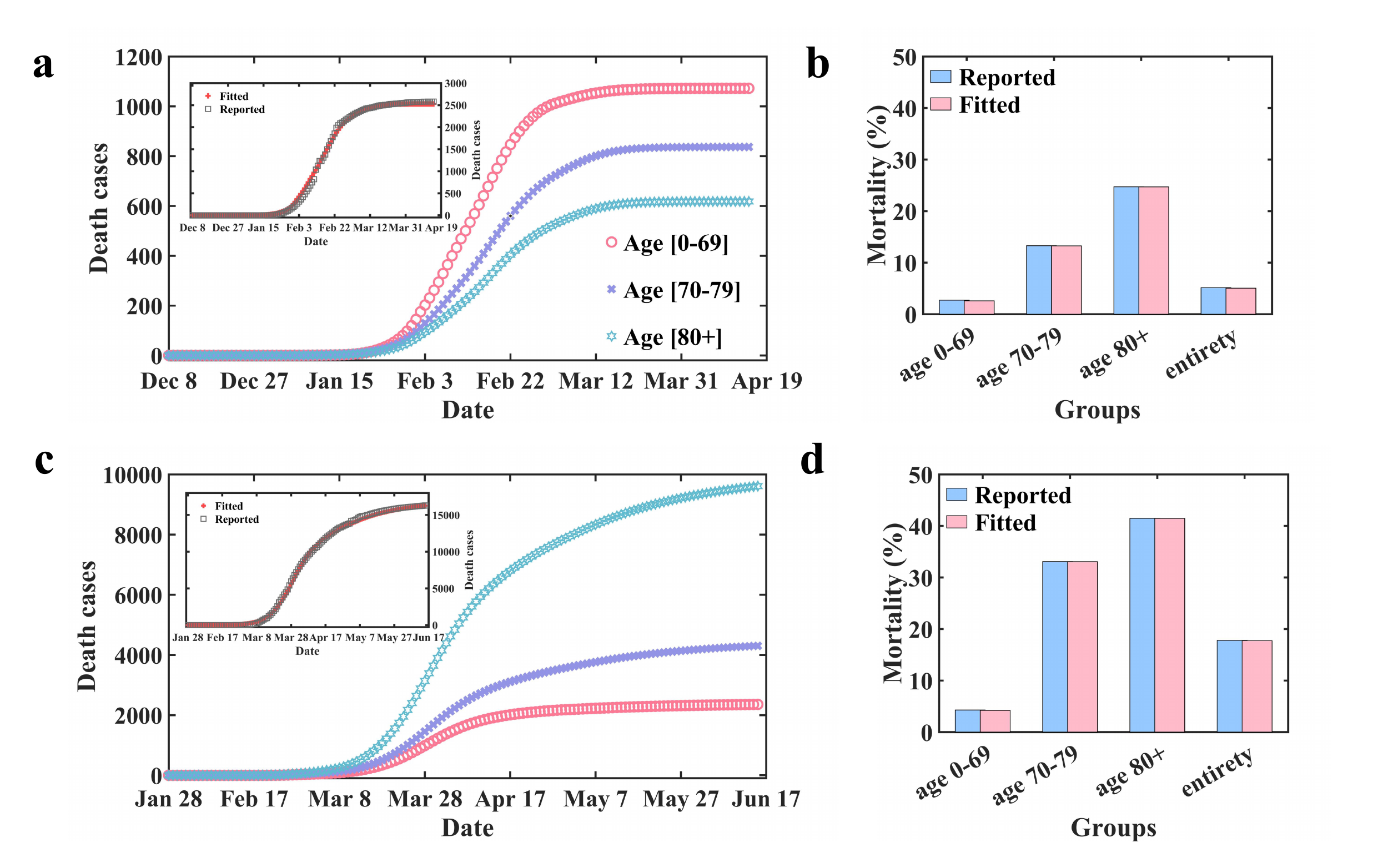}
\caption{ {\em Evolution of death toll with time for different age groups.}
(a) Simulated evolution trend of the death toll of the three age groups in
Wuhan; (b) the actual mortality rate of each age group in Wuhan in comparison
with model prediction; (c,d) the corresponding results for Lombardy.
The insets in (a) and (c) display the actual daily death toll and model
predictions for Wuhan and Lombardy, respectively. In panels (b) and (d), the
light blue and pink columns represent the actual and simulated mortality rate
of each group, respectively.}
\label{fig:YF}
\end{figure}

Individuals in any age group have the possibility of getting infected by
COVID-19 with certain death risk~\cite{label19}. Currently available data
indicate that, in most countries, the risk of hospitalization and death
increases with age~\cite{label20,label21}. A direct manifestation is that
the fraction of hospitalized patients and ICU mortality vary among the age
groups.

For Wuhan, the available data divide patients into three groups:
[0-69] years old, [70-79] years old, and [80+] years old. A similar division
scheme applies to Lombardy. We use a linear regression to obtain the estimates
of the fractions of various state transitions for each age group:
$P_{M\rightarrow F}$, $P_{G\rightarrow W}$, $P_{F\rightarrow C}$
and $P_{U\rightarrow D}$. (SN4 provides more details.)

We input the parameters associated with each age group into the model and
calculate the death tolls as a function of time for Wuhan and Lombardy.
Figure~\ref{fig:YF} shows that the mortalities obtained from the simulation
agree well with those of the actual data. Figures~\ref{fig:YF}(a) and
\ref{fig:YF}(c) show that the death toll of patients over 80 years old in
Wuhan is the lowest among the age groups, while it is the highest in Lombardy:
about 59\%. There are two possible reasons for this difference. First, the
aging populations in the two regions are different: the fraction of people
over 65 years old is 14.06\% in Wuhan and 20\% in
Lombardy~\cite{label22,label23}. The elderly have weak autoimmunity and most
of them have preexisting, underlying diseases, resulting in an excessively
high death rate in Lombardy. The second reason is that more extensive tests
were conducted in China (near 100\%), enabling more young people with mild
symptoms to be detected and included in the statistics. In contrast, as of
April 29, 2020, there were about 1910761 people in Italy who had been tested,
of whom 203591 were diagnosed with COVID-19 - about 11\% only~\cite{label24}.
Nonetheless, Figs.~\ref{fig:YF}(b) and \ref{fig:YF}(d) show that the mortality
rate of patients over 70 years old is higher than that of patients under 70
for both Wuhan and Lombardy. Especially, in Lombardy, the mortality rate of
patients over 80 years old reached 41\%.

When the population is divided into nine age groups, our model generates
essentially the same phenomenon that the risk of death increases with
age (SN5).

\subsubsection*{Group weighting strategy}

People of all ages have certain risk of being infected with COVID-19. Setting
priority of admission for certain age groups can reduce the number of deaths.
In Wuhan, the ICU resources were more scarce than GW resources, so making ICU
admission policy dependent on age is especially important for suppressing the
mortality rate. That is, the conventional FCFS admission policy is not suitable
for COVID-19 under limited ICU resources. To set the priority of admission for
different age groups, we shift the state age of each group by a priority
weight. We then change the admission order of different age groups with the
goal to find an optimal set of weights that minimizes the number of deaths.
In particular, we set the priority weight of the $i_{th}$ age group as $w_i$
and record $\tau_i$ as the state age of the this age group. After weighting,
the new state age $\tau_i^{'}$ becomes,
\begin{equation}
\tau_i^{'}=\tau_i+w_i.
\end{equation}
The state age of patients in the $i_{th}$ age group is increased by $w_i$
days. The patients, after incorporating the weighted state age, are admitted
according to FCFS.

To carry out the optimization procedure, we fix the priority weight of the
second age group, i.e., the [70, 80) age group, to be $w_2 = 30$. As the state
age of each group does not exceed 15, we implement different admission
strategies for the age groups by changing the value of $w_1$ and $w_3$
(see SN5 for a specific method).

\begin{figure} [ht!]
\includegraphics[width=0.9\linewidth]{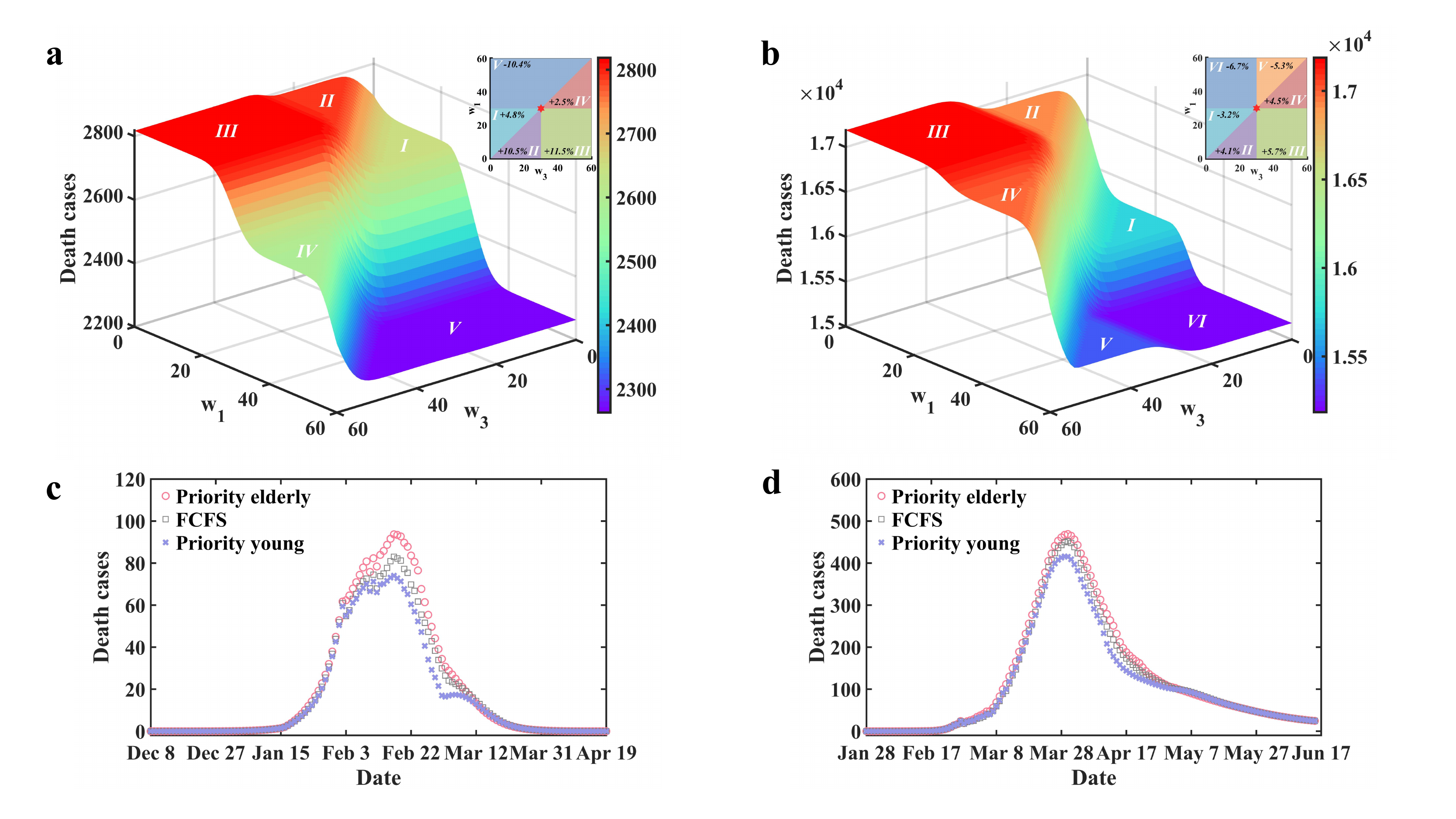}
\caption{ {\em Dependence of death toll on weights.}
(a,b) Model generated death toll and its dependence on $w_1$ and $w_3$ for
fixed $w_2 = 30$ for Wuhan and Lombardy, respectively, where the blue and
purple colors correspond to relatively fewer deaths. In the regions marked by
Roman numerals, the death toll changes slowly with weights. The insets show
the distribution of the number of deaths for different weight combinations,
where the numerical values represent the differences in the number of deaths
than that from the FCFS treatment strategy. (c,d) Daily number of new deaths
under different admission strategies for Wuhan and Lombardy, respectively,
where red circles, gray boxes, and blue-purple forks denote the results from
the three strategies: preferential treatment for elderly patients, FCFS, and
preference for young patients, respectively. Giving priority to young patients
results in the lowest daily death curve.}
\label{fig:JQ}
\end{figure}

Figure~\ref{fig:JQ} shows the effect of the state age of groups 1 and 3 on
the number of deaths when the weight of group 2 is fixed at constant (30).
The larger the state age, the higher priority of admission is. When adjusting
the weights of the first and third groups, their correspond state ages will
vary and so the admission order of each age group will change accordingly.
Different orders of treatment will lead to different death tolls.
Figure~\ref{fig:JQ} reveals, in the parameter plane of $(w_1,w_3)$, five
distinct death toll regions for Wuhan. For Lombardy, there are six such
regions, as shown in Fig.~\ref{fig:JQ}(b). The weight range in region V in
Fig.~\ref{fig:JQ}(a) corresponds to the ranges in regions V and VI in
Fig.~\ref{fig:JQ}(b), due to the relatively small fraction of patients in the
third group in Wuhan. When patients in the first group are given the priority,
the treatment order of the patients in the third group has little impact on
the total number of deaths. While the patients in the third group in Lombardy
have a large proportion, giving the priority to the first group will have a
significant impact. In this case, the order of treatment for patients in the
third group will affect the total number of deaths.

In Fig.~\ref{fig:JQ}, the number of deaths in Wuhan and Lombardy in parameter
region I increases by 4.8\% and decreases by 3.2\%, respectively, in comparison
with the death tolls from the FCFS treatment strategy, for $w_2 > w_1 > w_3$.
Area II in Fig.~\ref{fig:JQ} shows the number of deaths for $w_2 > w_3 > w_1$.
Comparing with the FCFS strategy, the number of deaths increases by 10.5\% and
4.1\% for Wuhan and Lombardy, respectively. In area III in Fig.~\ref{fig:JQ},
the weighting order is $w_3 > w_2 > w_1$ and the largest death tolls are
observed in Wuhan and Lombardy, as characterized by an increase of 11\% and
5.7\%, respectively, in comparison with the FCFS case. In area IV in
Fig.~\ref{fig:JQ} where the order of priority is $w_3 > w_1 > w_2$, the death
tolls in Wuhan and Lombardy increase by 2.5\% and 4.5\%, respectively. In
area V in Fig.~\ref{fig:JQ}(a), we have $w_1 > w_3 > w_2$, and the number of
deaths in Wuhan is the lowest as represented by a decrease of 10.4\% relative
to that associated with the FCFS strategy. In the same area V for Lombardy,
the corresponding decrease in the death toll is by 5.3\%, as shown in
Fig.~\ref{fig:JQ}(b). In area VI in Fig.~\ref{fig:JQ}(b), the priority
ordering is $w_1 > w_2 > w_3$ and there is a decrease of 6.7\% in the death
toll. The implication of these results is that giving the treatment priority
to young patients will reduce the number of deaths but giving priority to
elderly patients will increase the death toll. Due to the resource limit,
this strategy is intuitively reasonable and has in fact been commonly practiced
in many hospitals and healthcare facilities. Our results provide a validation
at a quantitative level.

Figures~\ref{fig:JQ}(c) and \ref{fig:JQ}(d) demonstrate the evolution of new
deaths over time in Wuhan and Lombardy, respectively, under different admission
strategies. In both cases, there is no significant difference in the number
of deaths in the early stages of the epidemic under three treatment strategies.
When the epidemic has lasted for a period of time and the death toll increases
sharply, giving priority to young patients can significantly reduce the number
of deaths. At the end of the epidemic, again the differences among the three
treatment strategies diminish. Taken together, these results verify that,
under limited medical resources, during the rapidly increasing phase of COVID-19
infection, admitting and treating young patients are necessary to reduce the
final death toll.

\subsubsection*{Optimization efficiency}

When the ICU resources are in a serious shortage, medical staff have
to face the hard choice of adopting the strategy of giving priority to
treating young patients with a higher survival probability. The strategy's
effectiveness notwithstanding, it has serious implications for medical ethics.
It is thus worth evaluating this admission/treatment strategy further. To this
end, we exploit our model to study more extensively the impacts of different
strategies for four countries: China, South Korea, Italy, and Spain. For each
country, we collect information about the numbers of GW and ICU beds per
100000 population and about the age distribution of patients diagnosed with
COVID-19. For example, in China and Italy, the numbers of ICU beds per
100000 individuals are 3.6 and 12.5, respectively. (A detailed display of
the information can be found in SN6). We define the time-dependent
optimization efficiency as
\begin{equation}
E(t)=\frac{V_Q(t)}{V_Y(t)},
\end{equation}
where $V_Q(t)$ and $V_Y(t)$, respectively, are the numbers of new deaths per
day with the FCFS and the age-selective strategy. In the asymptotic time limit
$t\rightarrow \infty$ when the system has reached a steady state, $E(\infty)$
characterizes the final optimization efficiency. Since information about the
average state transition fractions for patients in all four countries is
not available, we use the model parameter values for Wuhan and Lombard to
calculate the value of $E(\infty)$ for four countries versus the average daily
patient size.

\begin{figure} [ht!]
\includegraphics[width=0.8\linewidth]{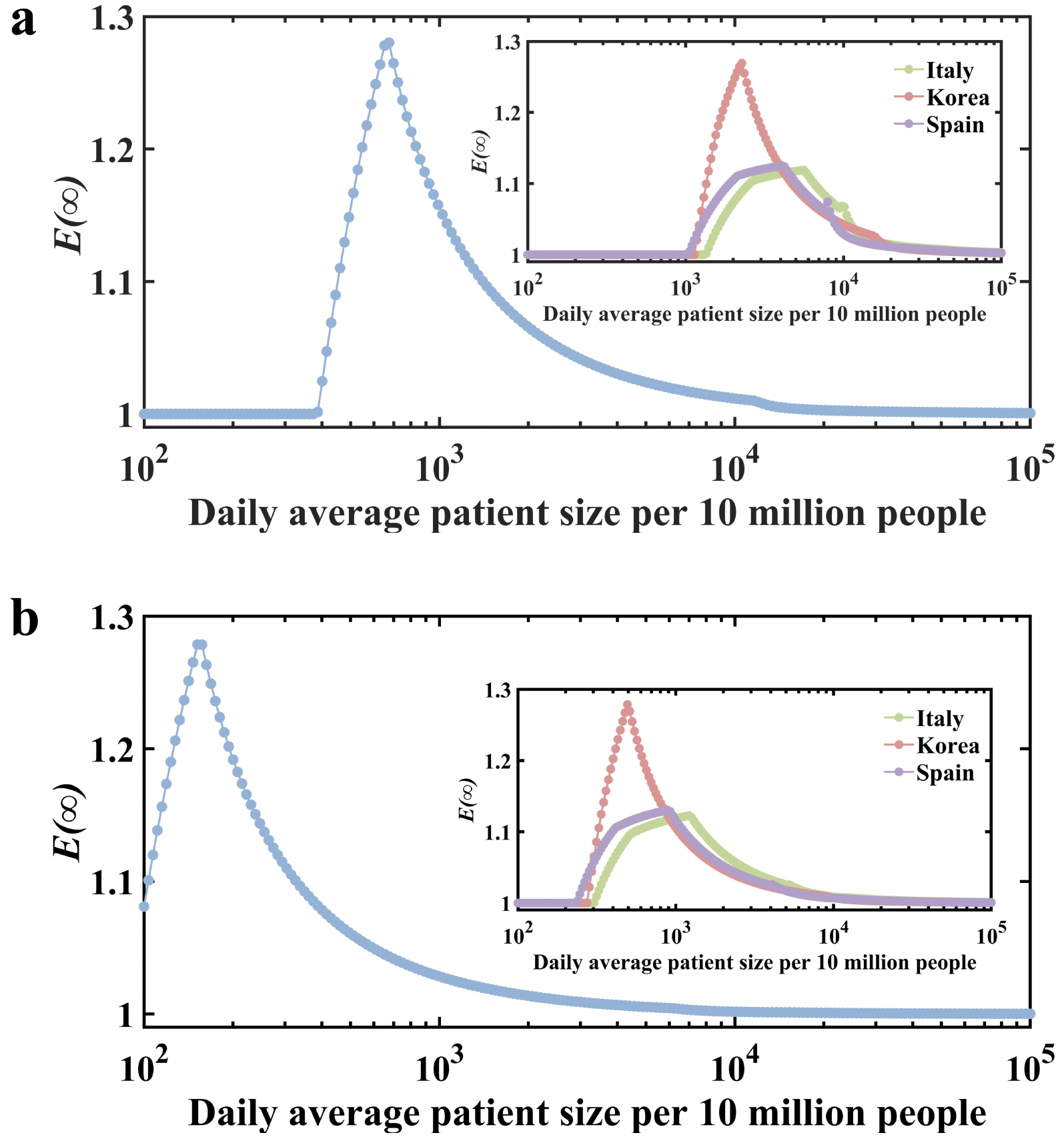}
\caption{ {\em Asymptotic optimization efficiency in China, Korea, Italy,
and Spain.} Shown is the steady-state efficiency $E(\infty)$ versus the cases
per ten million people. (a) The result for China based on the parameter values
from Wuhan. Inset shows the corresponding results for Italy, South Korea,
and Spain. (b) The result for China based on the parameter values from
Lombardy and the results for the other three countries are displayed in
inset.}
\label{fig:YH}
\end{figure}

Based on the per capita medical resource data of the four countries as well
as the age distribution data of confirmed patients in each country, making
use of the relevant model parameter values for Wuhan and Lombard, we calculate
$E(\infty)$ of the four countries versus the daily average patient size. We
assume that up to 50\% of the medical resources can be deployed to treating
patients with COVID-19. As shown in Fig.~\ref{fig:YH}, when the number of
patients is below 100, the value of $E(\infty)$ for the four countries is
equal to one. In this case, there is no need to implement the age-selective
admission strategy. When the number of patients exceeds a critical value,
$E(\infty)$ for all four countries increases rapidly as the number of patients
increases through the critical value and then reaches a peak value. In fact,
for a wide range of variation of the patient size, the values of $E(\infty)$
are high. Note that the peak values of $E(\infty)$ for China and South Korea
are slightly below 1.3, but those for Italy and Spain are lower than 1.2 with
a relatively slow increasing rate to reach the peak value. A plausible reason
lies in the difference in the age distribution of patients diagnosed with
COVID-19 in the eastern and western countries. In particular, in China and
South Korea, less than 12\% of the patients are over 70 years old, while
those in Italy and Spain account for 37\% and 34\%, respectively (SN6).
The value of $E(\infty)$ is highly correlated with the availability of the
ICU beds. For $E(\infty)=1$, ICU beds are fully sufficient, but $E(\infty)>1$
signifies a shortage of ICU beds. Figure~\ref{fig:YH}(a) reveals that, with
the per capita medical resources of South Korea, Italy, and Spain, the need
for ICU of 1,000 cases per tens of millions of people per day can be met for
a long time, while the capacity of China may be less than half of the capacity
of those countries. Indeed, the number of ICU beds per capita in China is 3.6
per 100000 persons, while the corresponding numbers in Italy, South Korea, and
Spain are 12.5, 10.6, and 9.7, respectively (SN 6). We conclude that, while
the age-selective admission strategy is more effective in countries with
a younger population structure, countries with less per capita ICU medical
resources need to implement this strategy, particularly in the early stage
of the pandemic when the number of patients is relatively small.

\section*{Discussion}

With the resurgence of COVID-19 cases in many countries and regions, a severe
shortage of medical resources for treating the disease is inevitable,
potentially leading to a significant death toll. To study the impact of
limited medical resources on the patient mortality at a quantitative level
can provide insights into developing optimal resource allocation schemes to
reduce the number of deaths. Based on the COVID-19 data, medical resources
and other relevant information, we have developed an admission treatment model
subject to limited medical resources, which enables a quantitative and
systematic assessment of the mortality rate associated with various resource
allocation scenarios. Using empirical data from Wuhan and Lombardy, we
validate the model by demonstrating that it can reproduce accurately the
evolution of daily death toll in both places. The validated model is then
used to assess the impact of different scenarios of medical deployment
(including deployment time and resource investment) on the number of deaths.
In general, the ICU resources have a significant impact on the mortality
rate, and it is intuitively evident that deploying the ICU resources earlier
and/or augmenting them will reduce the death toll. A virtue of our model is
that it enables an accurate quantification of such intuitive expectations.
For example, we find that, if the deployment of ICU resources had been made
one week earlier or if they had been doubled, the death toll in Wuhan would
have been reduced by 5\% or 13\%, and that in Lombardy by 3\% and 14\%,
respectively. If the number of ICU beds had been tripled, the death toll in
both Wuhan and Lombardy would have been reduced by 21\%.

Our model enables an accurate prediction of the COVID-19 death toll for any
combination of the two parameters: the amount of advance or delay in the
timing of ICU deployment and the enhancement factor of the ICU resources.
Our model asserts that simultaneously advancing the timing of ICU resource
deployment and enhancing the ICU resources will reduce the death toll
significantly. For example, if the ICU resources had been augmented by a
factor of 2.5 and deployed two weeks earlier than the actual date, the final
death toll would have been reduced by 29\% in Wuhan. In Lombardy, if the
number of ICU beds had been three times the actual number and they had been
placed into service one week earlier, the final death toll would have been
reduced by 22\% and the ICU ward overload days would have been suppressed to
less than one week.

Likewise, our model is fully capable of assessing the impacts of GW resources
on COVID-19 mortalities. For example, if the GW resources in Wuhan had been
deployed one week in advance or if the GW resources had been doubled, the death
toll would have been reduced by 10\% comparing with the actual toll. However,
the model predicts that deploying GW resources earlier than one week or
augmenting them further would have little effect on suppressing the death toll
further. For Lombardy, the GW resources play an even less role in reducing
the death toll: augmenting them or deploying them earlier would reduce the
death toll by about 1\% at maximum. This indicates that the GW resources in
Lombardy were sufficient. This should be contrasted with the scenario of
significantly enhancing the ICU resources, e.g., by a factor of four, where
the death toll would be reduced by 26\% and 22\% for Wuhan and Lombardy,
respectively. The implication is that, in comparison with the GW resources,
the actual ICU resources in Wuhan and Lombardy were far from sufficient.

The morbidity and mortality of patients infected with COVID-19 increase with
age. We have studied the role of age selective admission and treatment
strategies in the death toll. In general, giving admission priorities to
patients in different age groups would result in different mortality rate.
In particular, if priorities had been given to the younger age groups, the
number of deaths in Wuhan and Lombardy would have been reduced by 10.4\% and
6.7\%, respectively, in comparison with that from the normal FCFS strategy.
In contrast, if priority had been given to elderly patients, the number of
deaths would have increased by 11.5\% and 5.7\% for the two places,
respectively.

The optimal admission/treatment strategies also depend on the scale of COVID-19
outbreak, the age structure of patients in the general population, and the per
capita medical resources. We have quantified these effects by defining the
steady-state optimization efficiency $E(\infty)$ and calculated this quantity
for four countries: China, South Korea, Italy, and Spain. One finding is that,
when the number of patients exceeds a threshold value, the values of
$E(\infty)$ for the four countries first rise rapidly with the increase of
the epidemic scale and then reaches a peak value. Remarkably, $E(\infty)$
can be maintained at a high value in a wide range of the epidemic scale.
The peak values of Italy and Spain are both lower than 1.2, and the rising
rate is slower. However, the peak values of China and South Korea are both
close to 1.3. This discrepancy between the two eastern Asian and two western
countries can be explained by the age distribution of the patients diagnosed
with COVID-19: in China and South Korea, less than 12\% of the diagnosed
patients are over 70 years old, while those in Italy and Spain account for
about 35\%. The per capita medical resources of South Korea, Italy and Spain
can meet the long term demand of 1000 cases per tens of million people per
day, while the caring capacity of China may be less than half of that of the
these countries. This is due to the small number of ICU beds per capita in
China, which is 3.6 per 100000 persons, compared with 12.5, 10.6, and 9.7 in
Italy, South Korea, and Spain, respectively. We conclude that the age-selective
admission strategy is more effective in countries with a younger age structure,
while the countries with less per capita ICU medical resources may need to
implement this admission strategy in the early stage of the epidemic when
the number of patients is relatively small.

Our model provides a general evaluation framework for assessing, at a
quantitative level, the necessary medical resource deployment and admission
strategy. It can be used to predict and articulate, under limited medical
resources, optimal scenarios with respect to resource deployment and hospital
admission/treatment strategies to minimize the death toll for future outbreaks
of infectious diseases.

\section*{Data Availability}

All relevant data are available from the authors upon request.

\section*{Code Availability}

All relevant computer codes are available from the authors upon request.

\section*{Acknowledgments}

We thank Zhai Zhengmeng and Long Yongshang for correcting the procedure,
Kang Jie for collecting data, and Lin Zhaohua and Han Lilei for discussing
the experimental design. This work was supported by the National Natural
Science Foundation of China (Grant Nos.~11975099, 11675056, 61802321 and
11835003), the Natural Science Foundation of Shanghai (Grant No.~18ZR1412200),
and the Science and Technology Commission of Shanghai Municipality
(Grant No.~14DZ2260800). YCL would like to acknowledge support from the
Vannevar Bush Faculty Fellowship program sponsored by the Basic Research
Office of the Assistant Secretary of Defense for Research and Engineering
and funded by the Office of Naval Research through Grant No.~N00014-16-1-2828.

\section*{Author Contributions}
M.T. designed research; Y.-Q.Z., L.Z. performed research; Y.-Q.Z., L.Z., M.T., Y.L., Z.L., and Y.-C.L. analyzed data; Y.-Q.Z., L.Z., M.T., Z.L., and Y.-C.L. wrote the paper.

\section*{Competing Interests}

The authors declare no competing interests.

\section*{Correspondence}

To whom correspondence should be addressed. E-mail: tangminghan007@gmail.com; Ying-Cheng.Lai@asu.edu

%\bibliographystyle{naturemag}
%\bibliography{COVID_19}
%\end{document}

%\bibliographystyle{abbrv} %%% reference 1. 2.
%\makeatletter
%\renewcommand\@biblabel[1]{[#1].}
%\makeatother

%\linespread{1.05}
%\begin{document}
\begin{titlepage}

\begin{center}
{\large Supplementary Information for}
\end{center}

\begin{center}
	{\large\bf Quantitative assessment of the effects of resource optimization and ICU admission policy on COVID-19 mortalities}
\end{center}
\begin{center}
{ Ying-Qi Zeng, Lang Zeng, Ming Tang$^*$, Ying Liu, Zong-Hua Liu, and Ying-Cheng Lai$^*$}
\end{center}

%\begin{center}
%	Corresponding author: Ming Tang (tangminghan007@gmail.com), Ying-Cheng Lai(Ying-Cheng.Lai@asu.edu)
%\end{center}
%\tableofcontents

\end{titlepage}

\section*{\large Supplementary Note 1: Equations governing non-Markovian
dynamical evolution of states}

Because of the non-Markovian nature of the dynamical evolution of the states,
it is convenient to describe the underlying process in terms of difference
equations over infinitesimal intervals in both time delay and time. In the
following, we derive, one by one, the difference equations for all the ten
states in our model.

\subsection*{\normalsize M state}

The evolution equation of the number of M-state patients with state age $\tau$
at time $t$ is
\begin{equation}
\begin{aligned}
M(\tau  + d\tau ;t + dt) &= [1 - {\omega _{M \rightarrow F}}(\tau )d\tau]\frac{{{P_{M \rightarrow F}}\int_\tau ^{+\infty}  {{f_{M \rightarrow F}}(\tau^{'})} d\tau^{'}}}{{{P_{M \rightarrow F}}\int_\tau ^{+\infty}  {{f_{M \rightarrow F}}(\tau^{'})} d\tau^{'} + (1 - {P_{M \rightarrow F}})\int_\tau ^{+\infty}  {{f_{M \rightarrow R}}(\tau^{'})} d\tau^{'}}}M(\tau ;t)\\
&+[1 - {\omega _{M \rightarrow R}}(\tau )d\tau]\frac{{(1-{P_{M \rightarrow F}})\int_\tau ^{+\infty}  {{f_{M \rightarrow R}}(\tau^{'})} d\tau^{'}}}{{{P_{M \rightarrow F}}\int_\tau ^{+\infty}  {{f_{M \rightarrow F}}(\tau^{'})} d\tau^{'} + (1 - {P_{M \rightarrow F}})\int_\tau ^{+\infty}  {{f_{M \rightarrow R}}(\tau^{'})} d\tau^{'}}}M(\tau ;t),
\end{aligned}
\end{equation}
where $M(\tau;t)$ is the density function of M-state patients with state age $\tau$
at time $t$, so the number of nodes in the M state with the state age in
the interval $(\tau,\tau+d\tau)$ is $M(\tau;t)d\tau$. A node in the M state
enters the F state (R state) at the conditional rate
$\omega_{M\to F}(\tau)$ [$\omega_{M\to R}(\tau)$], which are given by
\begin{eqnarray}
\nonumber
\omega_{M\to F}(\tau) & = & \frac{f_{M\to F}(\tau)}{\int_{\tau}^{+\infty}f_{M\to F}(\tau')d\tau'}, \\ \nonumber
\omega_{M\to R}(\tau) & = & \frac{f_{M\to R}(\tau)}{\int_{\tau}^{+\infty}f_{M\to R}(\tau')d\tau'},
\end{eqnarray}
where $f_{M \rightarrow F}(\tau)$ and $f_{M \rightarrow R}(\tau)$ are the
probability density functions of the time delay from the M state to the
F and R states, respectively, each representing the probability of a
state transition within the time interval $(t,t+dt)$. Since the M state can
transition to two different states: $M\rightarrow F$ and $M\rightarrow R$,
the two processes will compete with each other, so the remaining probabilities
of the state transitions at time $t$ are
\begin{equation}
\begin{aligned}
\frac{{{P_{M \rightarrow F}}\int_\tau ^{+\infty}  {{f_{M \rightarrow F}}(\tau^{'})} d\tau^{'}}}{{{P_{M \rightarrow F}}\int_\tau ^{+\infty}  {{f_{M \rightarrow F}}(\tau^{'})} d\tau^{'} + (1 - {P_{M \rightarrow F}})\int_\tau ^{+\infty}  {{f_{M \rightarrow R}}(\tau^{'})} d\tau^{'}}},
\end{aligned}
\end{equation}
and
\begin{equation}
\begin{aligned}
\frac{{(1-{P_{M \rightarrow F}})\int_\tau ^{+\infty}  {{f_{M \rightarrow R}}(\tau^{'})} d\tau^{'}}}{{{P_{M \rightarrow F}}\int_\tau ^{+\infty}  {{f_{M \rightarrow F}}(\tau^{'})} d\tau^{'} + (1 - {P_{M \rightarrow F}})\int_\tau ^{+\infty}  {{f_{M \rightarrow R}}(\tau^{'})} d\tau^{'}}},
\end{aligned}
\end{equation}
respectively.

The newly increased patients enter the M state. Suppose that $Z_{M}(t)$ new
patients are added within $(t,t+dt)$, we have
\begin{equation}
\begin{aligned}
M(0;t+dt)=Z_M(t).
\end{aligned}
\end{equation}
The state age of the newly added M-state patients is zero.

\subsection*{\normalsize F state}

The evolution equation of the number of F-state patients with state age $\tau$
at time $t$ is
\begin{equation}
\begin{aligned}
F(\tau  + d\tau ;t + dt) &= [1 - {\omega _{F \rightarrow C}}(\tau )d\tau]\frac{{{P_{F \rightarrow C}}\int_\tau ^{+\infty}  {{f_{F \rightarrow C}}(\tau^{'})} d\tau^{'}}}{{{P_{F \rightarrow C}}\int_\tau ^{+\infty}  {{f_{F \rightarrow C}}(\tau^{'})} d\tau^{'} + (1 - {P_{F \rightarrow C}})\int_\tau ^{+\infty}  {{f_{F \rightarrow M_X}}(\tau^{'})} d\tau^{'}}}F(\tau ;t)\\
&+[1 - {\omega _{F \rightarrow M_X}}(\tau )d\tau]\frac{{{(1-P_{F \rightarrow C})}\int_\tau ^{+\infty}  {{f_{F \rightarrow M_X}}(\tau^{'})} d\tau^{'}}}{{{P_{F \rightarrow C}}\int_\tau ^{+\infty}  {{f_{F \rightarrow C}}(\tau^{'})} d\tau^{'} + (1 - {P_{F \rightarrow C}})\int_\tau ^{+\infty}  {{f_{F \rightarrow M_X}}(\tau^{'})} d\tau^{'}}}F(\tau ;t),
\end{aligned}
\end{equation}
with notations similar in their meanings to those of the M-state equation.

After admission of patients in the C state, the remaining available GW
resources are $\Delta Q_{G}^{'}(t)$. There are two cases:
(i) the new resources available are sufficient to accommodate all F-state
patients; (ii) new available resources can only accept some or none of the
F-state patients. In these two cases, patients in the F state are admitted
according to the strategy of FCFS (first-come first-serve). The number of
F-state patients changing to the G state in $(t, t + dt)$ is denoted as
$\bigtriangleup_{F\rightarrow G}(t)$.\\
In the first case where the GW resources are sufficient to accommodate all
current F-states patients, if
\begin{displaymath}
\Delta Q_{G}^{'}(t)(t)\geq\int_{0}^{+\infty}F(\tau+d\tau;t+dt)d\tau,
\end{displaymath}
then all F-state individuals will change to the G state. We have
$F(\tau+d\tau;t+dt)=0$ for all $\tau$ $\in$ [0,$+\infty$), and
\begin{displaymath}
\bigtriangleup_{F\rightarrow G}(t)=\int_{0}^{+\infty}F(\tau+d\tau;t+dt)d\tau.
\end{displaymath}
In the second case where the GW resources can hold only some or none of the
current F-state patients, if
\begin{displaymath}
\Delta Q_{G}^{'}(t)(t)=\int_{\tau_c}^{+\infty}F(\tau+d\tau,t+dt)d\tau,
\end{displaymath}
then some F-state individuals will change to the G state. We have
$F(\tau+d\tau;t+dt)=0$ for $\tau$ $\in$ [$\tau_c$,$+\infty$), and
\begin{displaymath}
\bigtriangleup_{F\rightarrow G}(t)=\Delta Q_{G}^{'}(t)(t).
\end{displaymath}

When the patients in the M state enter into the F state, the initial state age
is set to be zero. The number of newly added F-state patients is given by
\begin{equation}
\begin{aligned}
F(0;t+dt)=\int_{0}^{+\infty}\omega_{M \to F}(\tau)\frac{P_{M\to F}\int_{\tau}^{+\infty}f_{M\to F}(\tau')d\tau'}{P_{M\to F}\int_{\tau}^{+\infty}f_{M\to F}(\tau')d\tau'+(1-P_{M\to F})\int_{\tau}^{+\infty}f_{M\to R}(\tau')d\tau'}M(\tau;t)d\tau .
\end{aligned}
\end{equation}

\subsection*{\normalsize C state}

The evolution equation of the number of C-state patients with state age $\tau$
at time $t$ is
\begin{equation}
\begin{aligned}
C(\tau+d\tau;t+dt)=[1-\omega_{C\to D}(\tau)d\tau]C(\tau;t).
\end{aligned}
\end{equation}
To get the number of C-state patients admitted to the hospital, we note
that there are two admission paths: (a) through ICU admission II to the
U state; (b) through GW admission I to the W state.

{\em ICU admission II path}.
After admission of W-state patients, the remaining available ICU resources
are $\Delta Q_{U}'(t)$. There are two cases: (i) the newly available
resources are sufficient to accommodate all the C-states patients and (ii) the
newly available resources are able to accommodate some or none of the C-state
patients. The number of C-state patients who switch to the U state in
$(t,t+dt)$ time interval is denoted as $\bigtriangleup_{C\rightarrow U}(t)$.

In case (i), if
\begin{displaymath}
\Delta Q_{U}'(t)\geq \int_{0}^{+\infty}C(\tau+d\tau;t+dt)d\tau,
\end{displaymath}
all C-state patients will change to the U state. We have $C(\tau+d\tau;t+dt)=0$
for $\tau \in$ [0,$+\infty$), and
\begin{displaymath}
\bigtriangleup_{C\rightarrow U}(t)=\int_{0}^{+\infty}C(\tau+d\tau;t+dt)d\tau.
\end{displaymath}
In case (ii), if
\begin{displaymath}
\Delta Q_{U}'(t)= \int_{\tau_c}^{+\infty}C(\tau+d\tau;t+dt)d\tau,
\end{displaymath}
and $0\leq \tau_{c}\leq +\infty$, then the C-state individuals whose state
ages are greater than $\tau_c$ will enter the U state. We have
$C(\tau+d\tau,t+dt)=0$ for $\tau \in$ [$\tau_c$,$+\infty$) and
\begin{displaymath}
\bigtriangleup_{C\rightarrow U}(t)=\Delta Q_{U}'(t).
\end{displaymath}
The remaining C-state patients can switch to the W state through the GW
admission I pathway.

{\em GW admission I path.}
Considering that the remaining C-state patients are in critical conditions,
when the ICU resources are unavailable to them, they will be admitted to GW
and change into the W state with the highest priority. The available resources
within the time interval $(t,t+dt)$ are denoted as $\Delta Q_{G}(t)$. There
are two cases: (i) the available GW resources are enough to accommodate all
the remaining C-state patients, and (ii) the available GW resources can
accommodate some or none of the remaining C-state patients. The number of
patients in the C state who change to the W state in $(t,t+dt)$ is denoted
as  $\bigtriangleup_{C\rightarrow W}(t)$.

In case (i), we have
\begin{displaymath}
\Delta Q_G(t)\geq\int_{0}^{+\infty}C(\tau+d\tau;t+dt)d\tau.
\end{displaymath}
The amount of the remaining available resources through F-state patients for
GW admission II is
\begin{displaymath}
\Delta Q'_G(t)=\Delta Q_G(t)-\int_{0}^{+\infty}C(\tau+d\tau;t+dt)d\tau.
\end{displaymath}
Individuals in the C state will change to the W state, so we have
$C(\tau+d\tau;t+dt)=0$ for $\tau \in$ [0,$+\infty$), and
\begin{displaymath}
\bigtriangleup_{C\rightarrow W}(t)=\int_{0}^{+\infty}C(\tau+d\tau;t+dt)d\tau.
\end{displaymath}

In case (ii), we have
\begin{displaymath}
\Delta Q_G(t)=\int_{\tau_c}^{+\infty}C(\tau+d\tau;t+dt)d\tau.
\end{displaymath}
The amount of the remaining available resources for GW admission II is
$\Delta Q'_G(t)=0$. The C-state individuals with state ages greater than
$\tau_c$ will enter the W state. We have $C(\tau+d\tau,t+dt)=0$ for
$\tau \in$ [$\tau_c$,$+\infty$) and
\begin{displaymath}
\bigtriangleup_{C\rightarrow W}(t)=\Delta Q_G(t).
\end{displaymath}

When the F-state patients enter the C state, the initial state age is set to
be zero. The number of newly added C-state patients is
\begin{equation}
\begin{aligned}
C(0;t+dt)=\int_{0}^{+\infty}\omega_{F\to C}(\tau)\frac{P_{F\to C}\int_{\tau}^{+\infty}f_{F\to C}(\tau)'d\tau'}{P_{F\to C}\int_{\tau}^{+\infty}f_{F\to C}(\tau)'d\tau'+[1-P_{F\to C}]\int_{\tau}^{+\infty}f_{F\to R}(\tau)'d\tau'}F(\tau;t)d\tau .
\end{aligned}
\end{equation}

\subsection{\normalsize G state}

The evolution equation of the number of G-state patients with state age $\tau$
at time $t$ is
\begin{equation}
\begin{aligned}
G(\tau+d\tau;t+dt)=[1-\omega_{G \to W}(\tau)d\tau]\frac{P_{G\to W}\int_{\tau}^{+\infty}f_{G\to W}(\tau')d\tau'}{P_{G\to W}\int_{\tau}^{+\infty}f_{G\to W}(\tau')d\tau'+(1-P_{G\to W})\int_{\tau}^{+\infty}f_{G\to M_X}(\tau')d\tau'}G(\tau;t)\\+[1-\omega_{G \to M_X}(\tau)d\tau]\frac{(1-P_{G\to W})\int_{\tau}^{+\infty}f_{G\to M_X}(\tau')d\tau'}{P_{G\to W}\int_{\tau}^{+\infty}f_{G\to W}(\tau')d\tau'+(1-P_{G\to W})\int_{\tau}^{+\infty}f_{G\to M_X}(\tau')d\tau'}G(\tau;t).
\end{aligned}
\end{equation}
The number of newly added G-state patients is
\begin{equation}
\begin{aligned}
G(0;t+dt)=\bigtriangleup_{F\rightarrow G}(t).
\end{aligned}
\end{equation}

\subsection{\normalsize W state}

The evolution equation of the number of W-state patients with state age $\tau$
at time $t$ is
\begin{equation}
\begin{aligned}
W(\tau+d\tau;t+dt)=(1-\omega_{W\to D}(\tau)d\tau)W(\tau;t).
\end{aligned}
\end{equation}
To have the number of W-state patients admitted to the hospital, we denote
the available ICU resources at time $t$ as $\Delta Q_U(t)$. There are two
cases: (i) there are sufficient resources to accommodate all current W-state
patients, and (ii) the available resources can accommodate some or none of
the W-state patients. The number of W-state patients who change to the U state
in $(t,t+dt)$ is $\bigtriangleup_{W\rightarrow U}(t)$.

In case (i), if
\begin{displaymath}
\Delta Q_U(t)\geq\int_{0}^{+\infty}W(\tau+d\tau;t+dt)d\tau,
\end{displaymath}
the amount of the remaining available resources is
\begin{displaymath}
\Delta Q'_U(t)=\Delta Q_U(t)-\int_{0}^{+\infty}W(\tau+d\tau;t+dt)d\tau.
\end{displaymath}
We have
\begin{equation}
\begin{aligned}
W(\tau+d\tau;t+dt)=0 \ for \ \tau \in [0,+\infty),\bigtriangleup_{W\rightarrow U}(t)=\int_{0}^{+\infty}W(\tau+d\tau;t+dt)d\tau.
\end{aligned}
\end{equation}
In case (ii), if
\begin{displaymath}
\Delta Q_U(t)=\int_{\tau_c}^{+\infty}W(\tau+d\tau,t+dt)d\tau
\end{displaymath}
and $0\leq \tau_c<+\infty$, the amount of the remaining available resources is
$\Delta Q'_U(t)=0$. We get
\begin{equation}
\begin{aligned}
W(\tau+d\tau;t+dt)=0 \ for \ \tau \in [\tau_c,+\infty), \bigtriangleup_{W\rightarrow U}(t)=\Delta Q_U(t).
\end{aligned}
\end{equation}
The sources of the W-state patients are the G-state patients whose conditions
have deteriorated and the C-state patients admitted through GW admission I.
The number of newly added W-state patients is
\begin{equation}
\begin{aligned}
W(0;t+dt)=\bigtriangleup_{C\rightarrow W}(t)+ \int_{0}^{+\infty}\omega_{G \to W}(\tau)\frac{P_{G\to W}\int_{\tau}^{\infty}f_{G\to W}(\tau')d\tau'}{P_{G\to W}\int_{\tau}^{\infty}f_{G\to W}(\tau')d\tau'+(1-P_{G\to W})\int_{\tau}^{\infty}f_{G\to M_X}(\tau')d\tau'}G(\tau;t)d\tau.
\end{aligned}
\end{equation}

\subsection*{\normalsize U state}

The evolution equation of the number of U-state patients with state age $\tau$
at time $t$ is
\begin{equation}
\begin{aligned}
U(\tau+d\tau;t+dt)=[1-\omega_{U \to D}(\tau)d\tau]\frac{P_{U\to D}\int_{\tau}^{+\infty}f_{U\to D}(\tau')d\tau'}{P_{U\to D}\int_{\tau}^{+\infty}f_{U\to D}(\tau')d\tau'+(1-P_{U\to D})\int_{\tau}^{+\infty}f_{U\to G_U}(\tau')d\tau'}U(\tau;t)\\+[1-\omega_{U \to G_U}(\tau)d\tau]\frac{(1-P_{U\to D})\int_{\tau}^{+\infty}f_{U\to G_U}(\tau')d\tau'}{P_{U\to D}\int_{\tau}^{+\infty}f_{U\to D}(\tau')d\tau'+(1-P_{U\to D})\int_{\tau}^{+\infty}f_{U\to G_U}(\tau')d\tau'}U(\tau;t).
\end{aligned}
\end{equation}
Patients in the U state are from the W state and the admitted C-state patients.
The number of newly added U-state patients is
\begin{equation}
\begin{aligned}
U(0;t+dt)=\bigtriangleup_{W\rightarrow U}(t)+\bigtriangleup_{C\rightarrow U}(t).
\end{aligned}
\end{equation}

\subsection*{\normalsize $G_{U}$ state}

The evolution equation of the number of $G_U$-state patients with state age
$\tau$ at time $t$ is
\begin{equation}
\begin{aligned}
G_U(\tau+d\tau;t+dt)=[1-\omega_{G_U\to M_X}(\tau)d\tau]G_U(\tau;t).
\end{aligned}
\end{equation}
The number of newly added $G_U$-state patients is
\begin{equation}
\begin{aligned}
G_U(0;t+dt)=\int_{0}^{+\infty}\frac{(1-P_{U\to D})\omega_{U\rightarrow G_U}(\tau)\int_{\tau}^{+\infty}f_{U\to G_U}(\tau')d\tau'}{P_{U\to D}\int_{\tau}^{+\infty}f_{U\to D}(\tau')d\tau'+(1-P_{U\to D})\int_{\tau}^{+\infty}f_{U\to G_U}(\tau')d\tau'}U(\tau,t)d\tau.
\end{aligned}
\end{equation}

\subsection*{\normalsize $M_X$ state}

The evolution equation of the number of $M_{X}$-state patients with state age
$\tau$ at time $t$ is
\begin{equation}
\begin{aligned}
M_X(\tau+d\tau;t+dt)=[1-\omega_{M_X \to R}(\tau)d\tau]M_X(\tau;t).
\end{aligned}
\end{equation}
The sources of $M_X$ state patients are three: patients from the F, G, and $G_U$
states. The number of newly added $M_X$-state patients is given by
\begin{equation}
\begin{aligned}
M_X(0,t+dt) &=\int_{0}^{+\infty}\omega_{G_U \to M_X}(\tau)G_U(\tau;t)d\tau \\&+\int_{0}^{+\infty}\frac{(1-P_{G\to W})\int_{\tau}^{+\infty}f_{G\to M_X}(\tau')d\tau'}{P_{G\to W}\int_{\tau}^{+\infty}f_{G\to W}(\tau')d\tau'+(1-P_{G\to W})\int_{\tau}^{+\infty}f_{G\to M_X}(\tau')d\tau'}\omega_{G\to M_X} G(\tau;t)d\tau \\&+\int_{0}^{+\infty}\frac{(1-P_{F\to C})\int_{\tau}^{+\infty}f_{F\to M_X}(\tau')d\tau'}{P_{F\to C}\int_{\tau}^{+\infty}f_{F\to C}(\tau')d\tau'+(1-P_{F\to C})\int_{\tau}^{+\infty}f_{F\to M_X}(\tau')d\tau'}\omega_{F\to M_X} F(\tau;t)d\tau.
\end{aligned}
\end{equation}

\subsection*{\normalsize R state}

The R-state patients come from the M and $M_X$ states. We have
\begin{equation}
\begin{aligned}
R(t+dt) &=R(t)+\int_{0}^{+\infty}\omega_{M_X\to R}M_X(\tau;t)d\tau\\&+\int_{0}^{+\infty}\frac{(1-P_{M\to F})\int_{\tau}^{+\infty}f_{M\to R}(\tau')d\tau'}{P_{M\to F}\int_{\tau}^{+\infty}f_{M\to F}(\tau')d\tau'+(1-P_{M\to F})\int_{\tau}^{+\infty}f_{M\to R}(\tau')d\tau'}\omega_{M\to R} M(\tau;t)d\tau.
\end{aligned}
\end{equation}

\subsection*{\normalsize D state}

The D-state patients come from C, W, and U states. We have
\begin{equation}
\begin{aligned}
D(t+dt) &=D(t)+\int_{0}^{+\infty}\omega_{C\to D}C(\tau;t)d\tau+\int_{0}^{+\infty}\omega_{W\to D}W(\tau;t)d\tau\\&+\int_{0}^{+\infty}\frac{P_{U\to D}\int_{\tau}^{+\infty}f_{U\to D}(\tau')d\tau'}{P_{U\to D}\int_{\tau}^{+\infty}f_{U\to D}(\tau')d\tau'+(1-P_{U\to D})\int_{\tau}^{+\infty}f_{U\to G_U}(\tau')d\tau'}\omega_{U \to D}(\tau)U(\tau;t)d\tau.
\end{aligned}
\end{equation}

\subsection*{\normalsize Calculation of available ICU resources $\bigtriangleup Q_U(t)$}

The available ICU resources are those that have been newly added, those
that are released when patients go from the U state to the D state or the $G_U$-state patients, and the resources consumed by the W-state and C-state
patients. We have
\begin{equation}
\begin{aligned}
\Delta Q_U(t+dt)&=\Delta Q_U(t)+Z_U(t)-\Delta_{W\rightarrow U}(t)-\Delta_{C\rightarrow U}(t) \\&+\int_{0}^{+\infty}\frac{P_{U\to D}\int_{\tau}^{+\infty}f_{U\to D}(\tau')d\tau'}{P_{U\to D}\int_{\tau}^{+\infty}f_{U\to D}(\tau')d\tau'+(1-P_{U\to D})\int_{\tau}^{+\infty}f_{U\to G_U}(\tau')d\tau'}\omega_{U \to D}(\tau)U(\tau;t)d\tau\\&+\int_{0}^{+\infty}\frac{(1-P_{U\to D})\int_{\tau}^{+\infty}f_{U\to G_U}(\tau')d\tau'}{P_{U\to D}\int_{\tau}^{+\infty}f_{U\to D}(\tau')d\tau'+(1-P_{U\to D})\int_{\tau}^{+\infty}f_{U\to G_U}(\tau')d\tau'}\omega_{U \to G_U}(\tau)U(\tau;t)d\tau.
\end{aligned}
\end{equation}

\subsection*{\normalsize Calculation of available GW resources $\Delta Q_G(t)$}

The available GW resources are those that have been newly added, those
released when patients switch from the G state to the $M_x$ state and from the
W state to the D state or the U state, and the resources consumed by the
C-state and F-state patients. We have
\begin{equation}
\begin{aligned}
\Delta Q_G(t+dt)&=\Delta Q_G(t)+Z_G(t)-\Delta_{C\rightarrow W}(t)-\Delta_{F\rightarrow G}(t)+\int_{0}^{+\infty}W(\tau;t)\omega_{W \to D}(\tau)d\tau\\&+\int_{0}^{+\infty}\frac{(1-P_{G\to W})\int_{\tau}^{+\infty}f_{G\to M_X}(\tau')d\tau'}{P_{G\to W}\int_{\tau}^{+\infty}f_{G\to W}(\tau')d\tau'+(1-P_{G\to W})\int_{\tau}^{+\infty}f_{G\to M_X}(\tau')d\tau'}\omega_{G \to M_X}(\tau)G(\tau;t)d\tau.
\end{aligned}
\end{equation}

\newpage

\section*{\large Supplementary Note 2: Parameter estimation}

\subsection*{\normalsize Average delay and fraction of state transition}

As shown in Tab.~\ref{tab:S1}, we assume that the time delay of patients
switching from state M to F follows the normal distribution with the average
of five days~\cite{label15}. The distributions of the time delays associated
with other state transitions are also assumed to be normal. The average time
delays from the U state to the D and $G_u$ states are set as seven
days~\cite{label16,label17} and eight days~\cite{label16}, respectively,
and that from the G state to the W state is three days~\cite{label15}.
In our model setting, the clinical symptoms of the F-state and G-state
patients are identical, and the difference lies in whether the patients
are admitted (i.e., occupying GW beds), and the same rule applies to the
C-state and W-state patients.

The average time delay from the F to the C state is two
days~\cite{label15,label25}, and that from the G and F states to the $M_X$
state is eight days~\cite{label10}. The average time delays from the W and
C states to the D state are assumed to be three days and one day, respectively.

As shown in Tab.~\ref{tab:S2}, we set the average mortality rate of ICU
patients as 61.5\%~\cite{label16,label26}. During the epidemic, due to
the different testing methods and conditions, the fraction of M-state patients
in the F state is different in different time periods. According to the
analysis of 32583 cases of laboratory-confirmed patients in Wuhan and
reconstruction of the epidemic trend~\cite{label27,label4}, we divide the
epidemic process in terms of the fraction of the transition from the M to
the F state into five stages. As shown in Table~\ref{tab:S3}, the time periods
and the transition fractions of the five stages are~\cite{label27}:
from 8 December 2019 to 10 January 2020 with the fraction
$|P|_{M\rightarrow F}(1) = 53.10\%$; from January 11 to January 22 with
$|P|_{M\rightarrow F}(2) = 35.10\%$; from January 23 to February 1 with
$|P|_{M\rightarrow F}(3) = 23.50\%$; from February 2 to February 16 with
$|P|_{M\rightarrow F}(4) =  15.90\%$; and after February 17 with
$|P|_{M\rightarrow F}(5) = 10.30\%$~\cite{label27}.

For Lombardy, the early epidemic can be divided into three
stages~\cite{label28}. We add a new time point: March 21, 2020, leading to
a division into four stages. After (including) this date, the intervention
measures in Lombardy and Italy as a whole reached maximum~\cite{label29}.
As shown in Table~\ref{tab:S4}, the time periods and the fractions associated
with the four stages are: from 28 January 2020 to 19 February 2020 with
$|P|_{M\rightarrow F}(1)=63\%$; from February 20 to February 25 with
$|P|_{M\rightarrow F}(2)=61\%$; from February 26 to March 20 with
$|P|_{M\rightarrow F}(3)= 56\%$~\cite{label28}; after March 21 with
$|P|_{M\rightarrow F}(4)= 25\%$.

By using the weighted least-squares method, we obtain the optimal estimates
of the parameters. In particular, for the Wuhan scenario, we obtain the optimal
set of parameters $\Xi=(|P|_{F\rightarrow C},|P|_{G\rightarrow W})$ by
minimizing the sum of the weighted difference squares between the reported
death curve~\cite{label30,label21} $D(t_i)$ and the model fitting value
$F(t_i,\Xi)$ $(i=0,1,\ldots, n-1)$, where
\begin{equation}
\begin{aligned}
	\widehat{\Xi}=\mbox{argmin} \sum_i w_{t_i}[F(t_i,\Xi)-D(t-i)]^{2}.
\end{aligned}
\end{equation}
To quantify the uncertainties of parameter estimation, we resort to the
general bootstrap method~\cite{label31,label32} and then obtain the 95\%
confidence interval of the estimated parameter values. As shown in
Tab.~\ref{tab:S2}, the optimal values and 95\% confidence interval of the
average transition fractions from G-state to W-state and from F-state to
C-state in Wuhan are $|P|_{F\rightarrow C}^{*}=48.3\%(41.06\%, 54.97\%)$ and
$|P|_{G\rightarrow W}^{*}=27.4\%(26.24\%, 28.51\%)$, respectively.

For Lombardy, the required optimal parameters are
$\Xi=(|P|_{M\rightarrow F}(4),|P|_{F\rightarrow C}, |P|_{G\rightarrow W})$.
As shown in Tabs.~\ref{tab:S2} and \ref{tab:S4}, the optimal values and the
95\% confidence intervals of the average transition fractions in Lombardy
are $|P|_{M\rightarrow F}^{*}=25\% \ (24.29\%, 25.65\%)$,
$|P|_{F\rightarrow C}^{*}=88.86\% \ (76.8\%, 100\%)$, and
$|P|_{G\rightarrow W}^{*}=63.77\% \ (62.9\%, 64.71\%)$, respectively.

\subsection*{\normalsize Incidence dates estimated from the confirmed data}

\begin{figure} [ht!]
\centering
\includegraphics[width=0.8\linewidth]{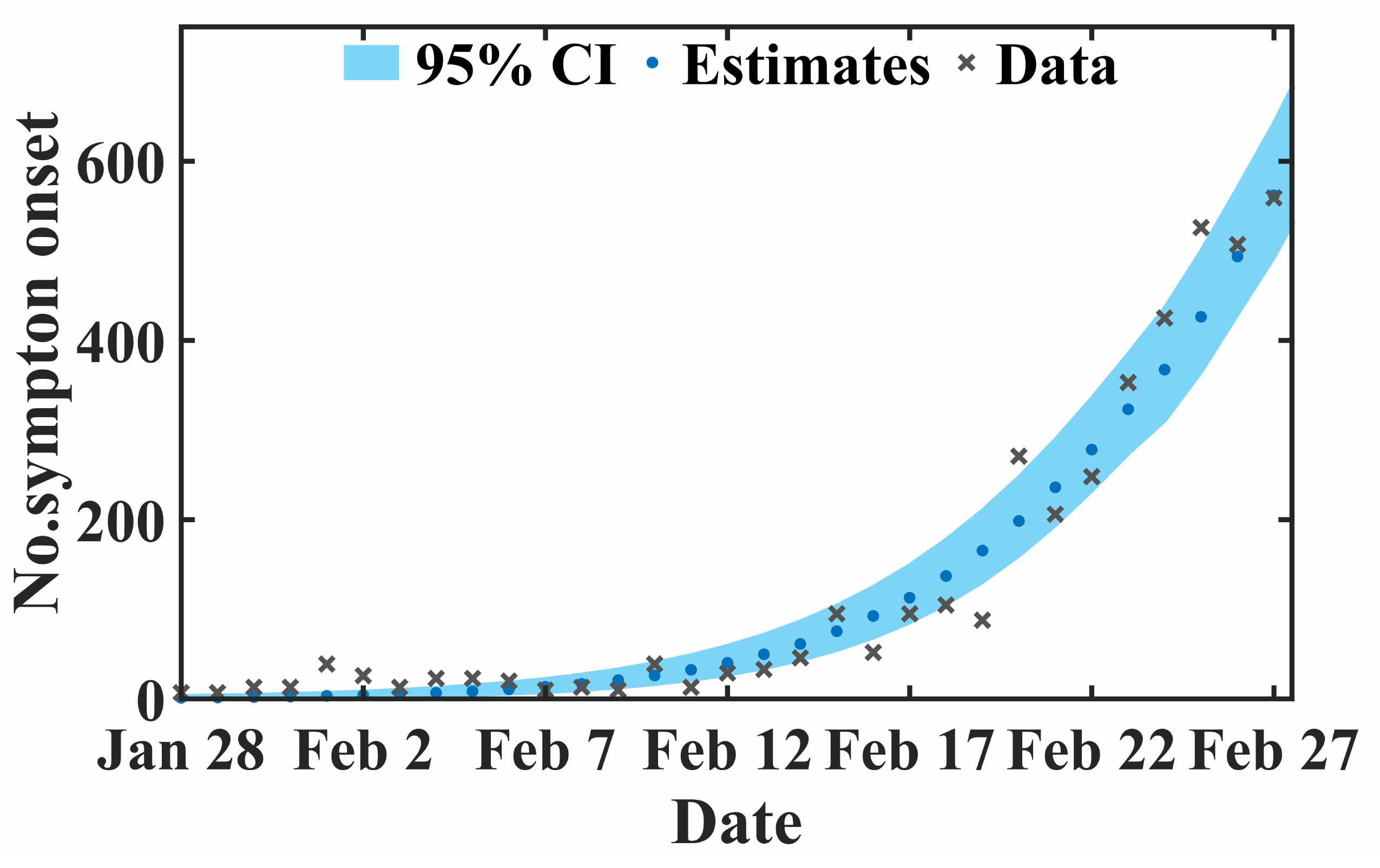}
\caption{ {\em Validation of our backtracking method}. The validation method
is based on the data of early laboratory-confirmed cases in Lombardy area
(January 28 - February 27, 2020). The incidence data of patients are
estimated from the laboratory reported confirmed cases and the distribution
of the time delay between the onset of the disease and the reported
confirmation. The blue dots and black crosses represent, respectively, the
incidence estimation curve and the retrospective survey data of
laboratory-confirmed cases in the literature~\cite{label33}. The light blue
shaded region represents the 95\% confidence interval of the estimation.}
\label{fig:S1}
\end{figure}

According to the average time delay from the onset date to the diagnosis
date of laboratory-confirmed cases in the literature~\cite{label27,label28},
we determine the distributions of state transition delay from onset to
diagnosis for Wuhan and Lombardy using a backtracking method based on
non-Markov processes that we have developed, where the incidence dates are
deduced from the confirmed cases. We designate a new state, the J state, to
denote that a patient has been confirmed. The dynamic process of J state to
M state is described by the following difference equations:
\begin{equation}
\begin{aligned}
J(\tau+d\tau;t-dt)&=[1-\omega_{J\rightarrow M}]J(\tau;t),\\
Z_M(t-dt)&=\int_{0}^{+\infty}J(\tau;t)\omega_{J\rightarrow M}(\tau)d\tau and\\
J(0;t-dt)&=Z_J(t),
\end{aligned}
\end{equation}
where $\omega_{J\rightarrow M}(\tau)$ is the conditional rate of transition
from J to M states with state age $\tau$, with the specific form
\begin{displaymath}
\omega_{J\to M}(\tau)=\frac{f_{J\to M}(\tau)}{\int_{\tau}^{+\infty}f_{J\to M}(\tau')d\tau'},
\end{displaymath}
with $f_{J \rightarrow M}(\tau)$ being the probability density function of
the transition time delay from J state to M state. For Wuhan and Lombardy, the state transition delays are assumed to follow the Gamma distribution with
the mean value of 9.5 and 7.3 days, respectively. The quantities $Z_M(t)$ and
$Z_J(t)$ are the numbers of new patients in the M and J states in the time
interval $(t,t+dt)$, respectively, where the former is our estimated value and
the latter is the newly reported, daily confirmed data. We validate the
accuracy of this method using data from early laboratory-confirmed cases in
Lombardy (January 28-February 27, 2020)~\cite{label33}, as shown in
Fig.~\ref{fig:S1}.

There are 50,008 laboratory/clinically confirmed cases in Wuhan, and 93,901 such
cases in Lombardy. We use the same backtracking method to trace the onset time
of the non-laboratory confirmed cases. The incidence dates of all confirmed
cases is April 15 for Wuhan and June 30 for Lombardy.

\subsection*{\normalsize Actual medical resource deployment}

In response to an emerging public health event, the government usually adopts
the policy of gradual deployment of medical resources to designated hospitals.
The dedicated medical resources in an area typically slowly increase with time.

As shown in Tabs.~\ref{tab:S5} and \ref{tab:S6}, we collect medical resource
deployment data dedicated to COVID-19 patients in Wuhan and Lombardy,
which include GW and ICU beds~\cite{label30,label34}. For Wuhan, we count the
number of open beds in critical hospitals. As the official reports do not
distinguish GW beds from ICU beds, we assume that ICU beds account for 4\% of
the total beds~\cite{label35}.

\newpage

\section*{\large Supplementary Note 3: Parameters of medical resource}

\subsection*{\normalsize Health care system stress metrics}

In addition to the final number of deaths, the load of local special medical
resources for COVID-19 patients is also an important indicator to measure the
stress of the medical system. We simulate and obtain the full-load working days
of GW and ICU in Wuhan and Lombardy during the recovery stage, denoted as
$O_G$ and $O_U$, respectively. When the GW or ICU is under full load, no new
patients can be admitted.

\begin{figure} [ht!]
\centering
\includegraphics[width=0.8\linewidth]{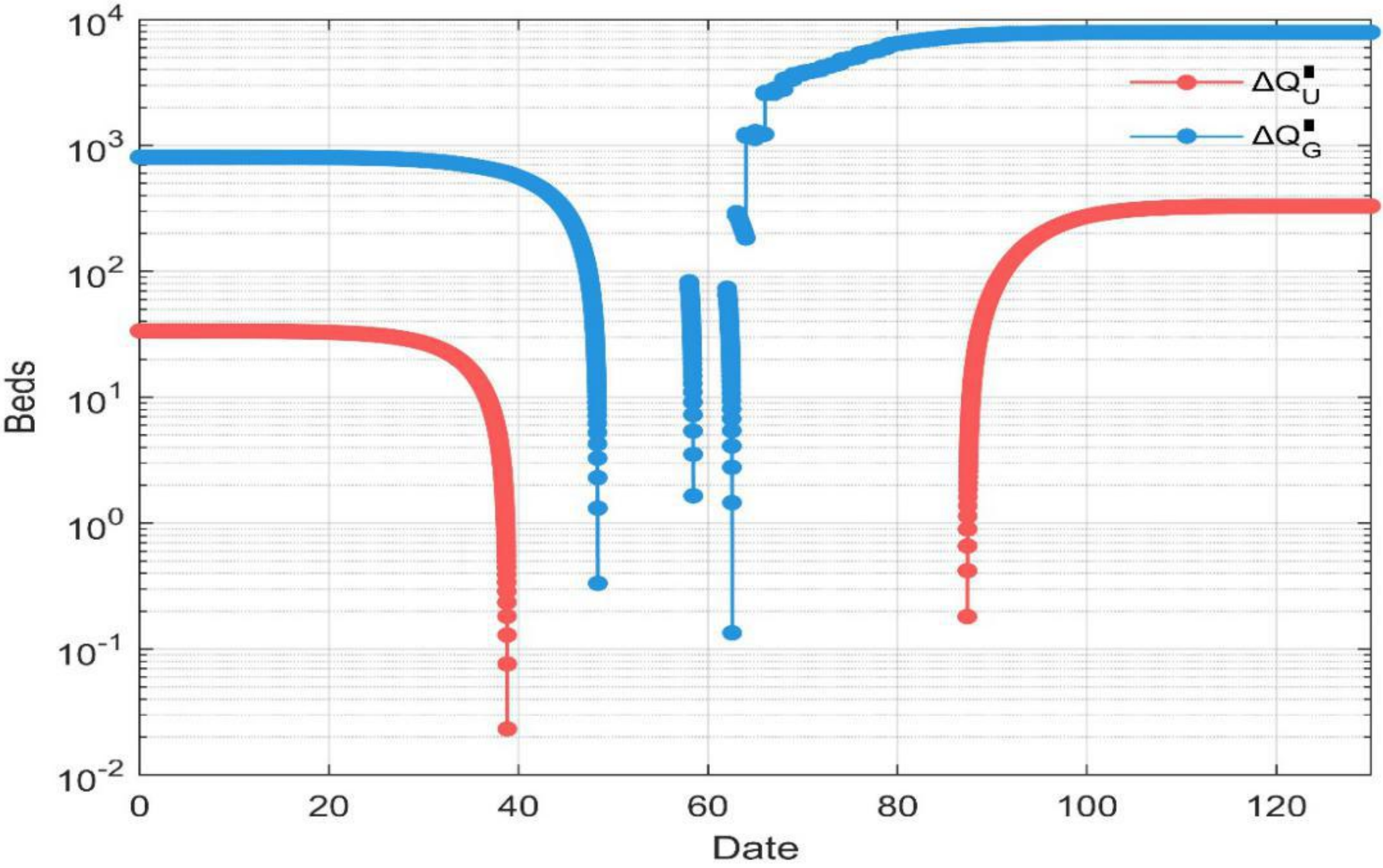}
\caption{ {\em Remaining resources of GW and ICU in Wuhan after two
admission processes.} The ordinate is on a logarithmic scale, and the red
and blue dots indicate that the amounts of remaining resources of Wuhan
ICU and GW at time $t$ are $\Delta Q_U(t)$ and $\Delta Q_G(t)$, respectively.
The missing areas of the red and blue dots represent the period of
$\Delta Q_U=0$ and $\Delta Q_G=0$, respectively.}
\label{fig:S2}
\end{figure}

In Fig.~\ref{fig:S2}, the remaining ICU resources after ICU admission I/II
at time $t$ is denoted as $\bigtriangleup Q_{U}(t)$, and the corresponding
quantity for GW is $\bigtriangleup Q_{G}(t)$, which are given by
\begin{equation}
\begin{aligned}
	O_U &=\int_{T_s}^{T_e}\delta[\triangle Q_{U}(t)]dt \ \ \mbox{and} \\
O_G &=\int_{T_s}^{T_e}\delta[\triangle Q_{G}(t)]dt,
\end{aligned}
\end{equation}
where $\delta(*)$ is the Dirac-$\delta$ function, the integration interval is
the whole recovery stage, $T_s$ and $T_e$ represent the starting and ending
time of the recovery stage, respectively. For example, $T_s$ in Wuhan is
December 8, 2019 and $T_e$ is April 16, 2020. The time interval is set to
be $dt=0.01$ day.

\subsection*{\normalsize Deployment plan of dedicated medical resources for COVID-19}

The deployment plan of local officially dedicated medical resources is
determined by two key factors: the deployment time DT and resource input RI,
where DT is the time for local authorities to start the deployment of the
dedicated medical resources and RI represents the financial and personnel
investment of the dedicated medical resources as characterized by the open
GW and ICU beds.

\begin{figure} [ht!]
\centering
\includegraphics[width=0.8\linewidth]{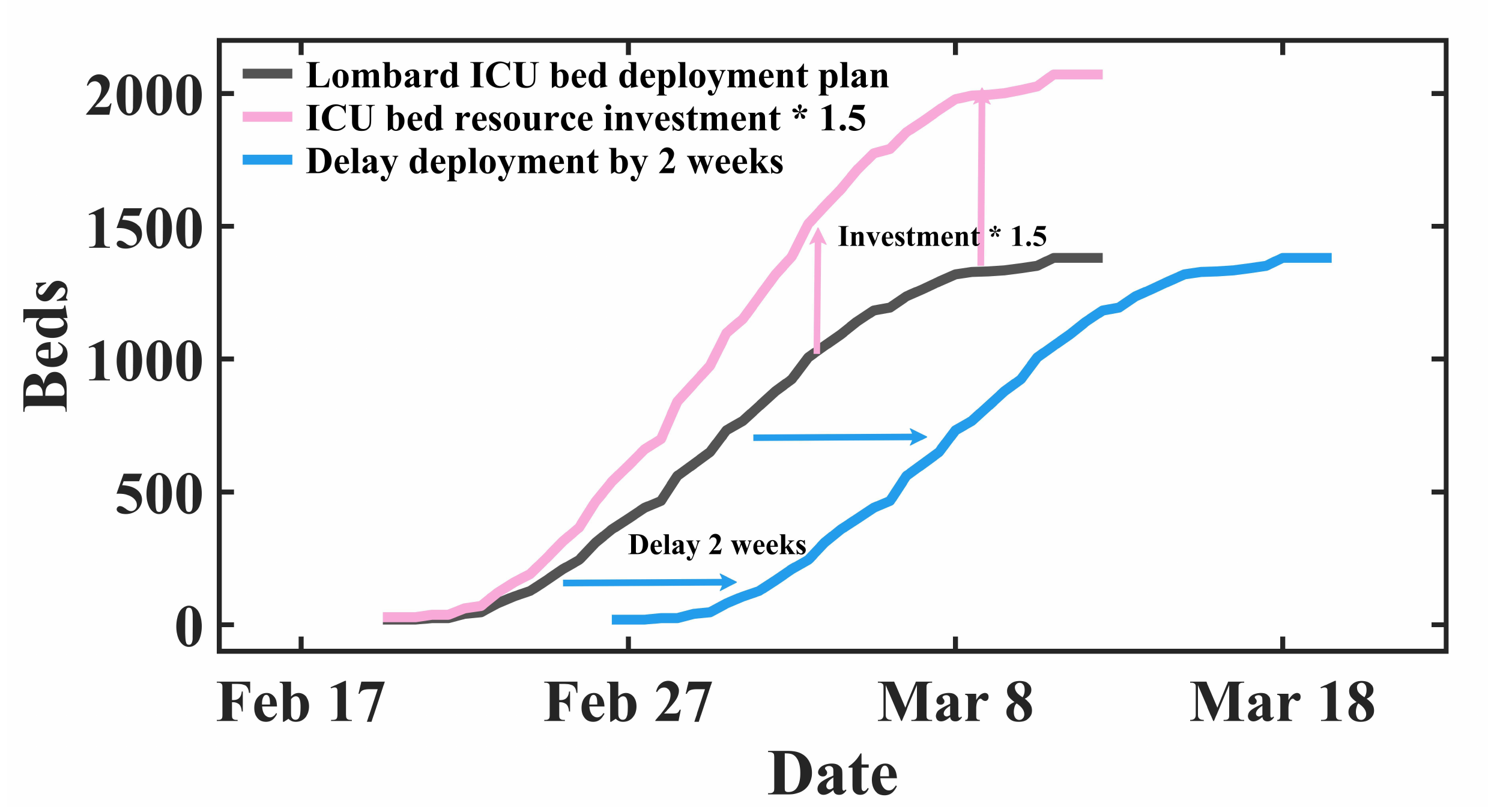}
\caption{ {\em Effects of varying ICU deployment plan in Lombardy, Italy.}
The black curve shows Lombardy's actual ICU medical resource deployment plan,
as displayed in Tab.~\ref{tab:S4}. The blue and red curves show the two cases
where Lombardy's government would delay the deployment of ICU resources by two
weeks ($\mbox{DT}=14$ and $\mbox{RI}=1$) and the ICU resource is expanded by
a factor of 1.5 ($\mbox{DT}=0$ and $\mbox{RI}=1.5$), respectively.}
\label{fig:S3}
\end{figure}

Figure~\ref{fig:S3} shows the effects of varying ICU deployment plan in
Lombardy, where the black curve represents Lombardy's current dedicated
ICU deployment plan. The actual deployment time serves as the benchmark
$\mbox{DT}=0$ and the actual resource input is $\mbox{RI}=1$. The left blue
arrow indicates the scenario where the Lombard authorities had delayed the
deployment time by two weeks ($\mbox{DT}=14$ and $\mbox{RI}=1$), and the
red up arrow corresponds to the scenario where the official resource input
had been increased by 50\% ($\mbox{DT}=0$ and $\mbox{RI}=1.5$).

\subsection*{\normalsize Effects of varying medical resource deployment on death toll}

\begin{figure} [ht!]
\centering
\includegraphics[width=0.8\linewidth]{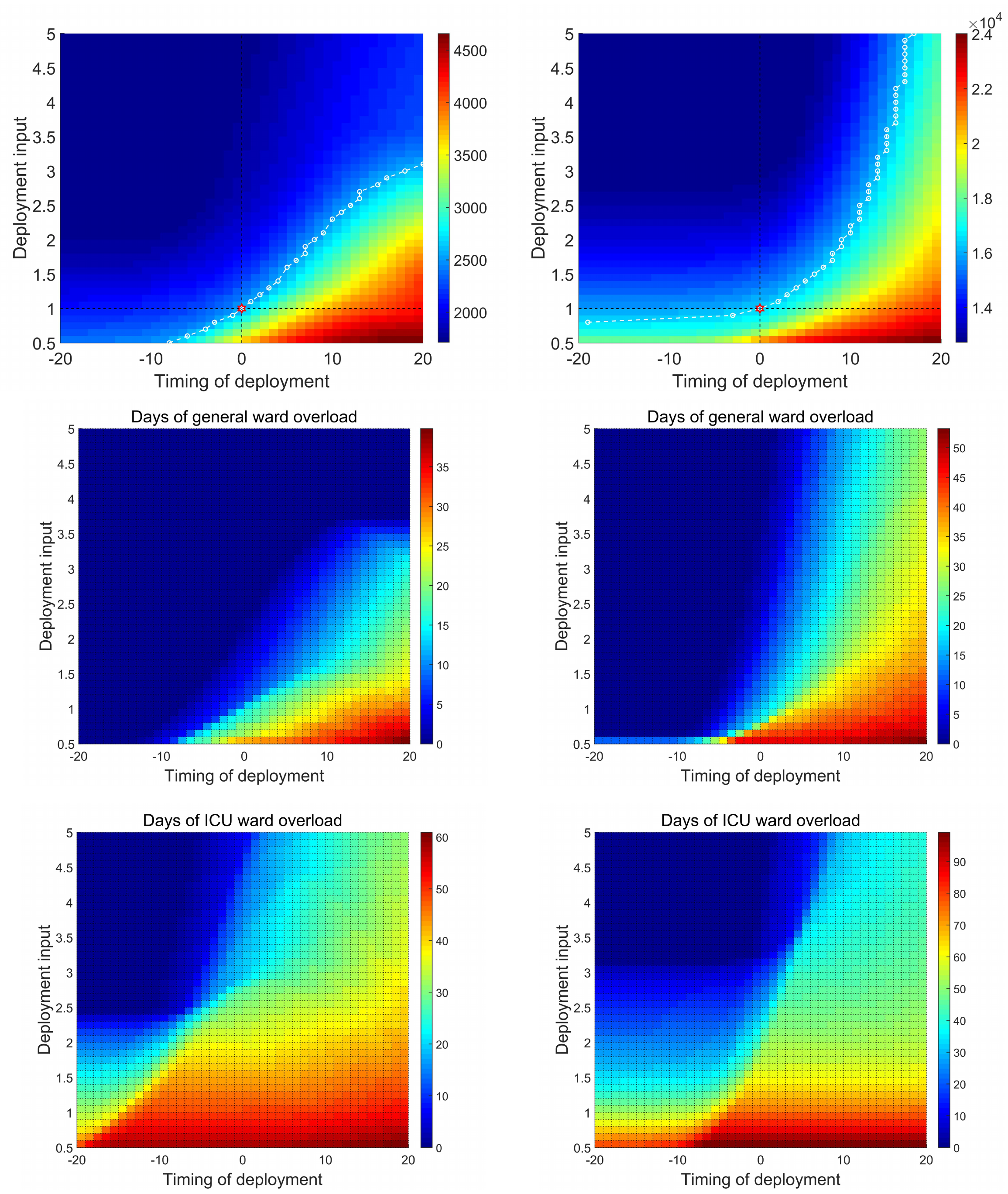}
\caption{ {\em Changes in the number of deaths and overload days for
simultaneous GW and ICU resource deployment for Wuhan and Lombardy}.
(a,b) Numbers of deaths in Wuhan and Lombardy, respectively. (c,d) GW ward
overload days in Wuhan and Lombardy, respectively. (e,f) ICU ward overload
days in Wuhan and Lombardy, respectively.}
\label{fig:S4}
\end{figure}

\begin{figure} [ht!]
\centering
\includegraphics[width=0.8\linewidth]{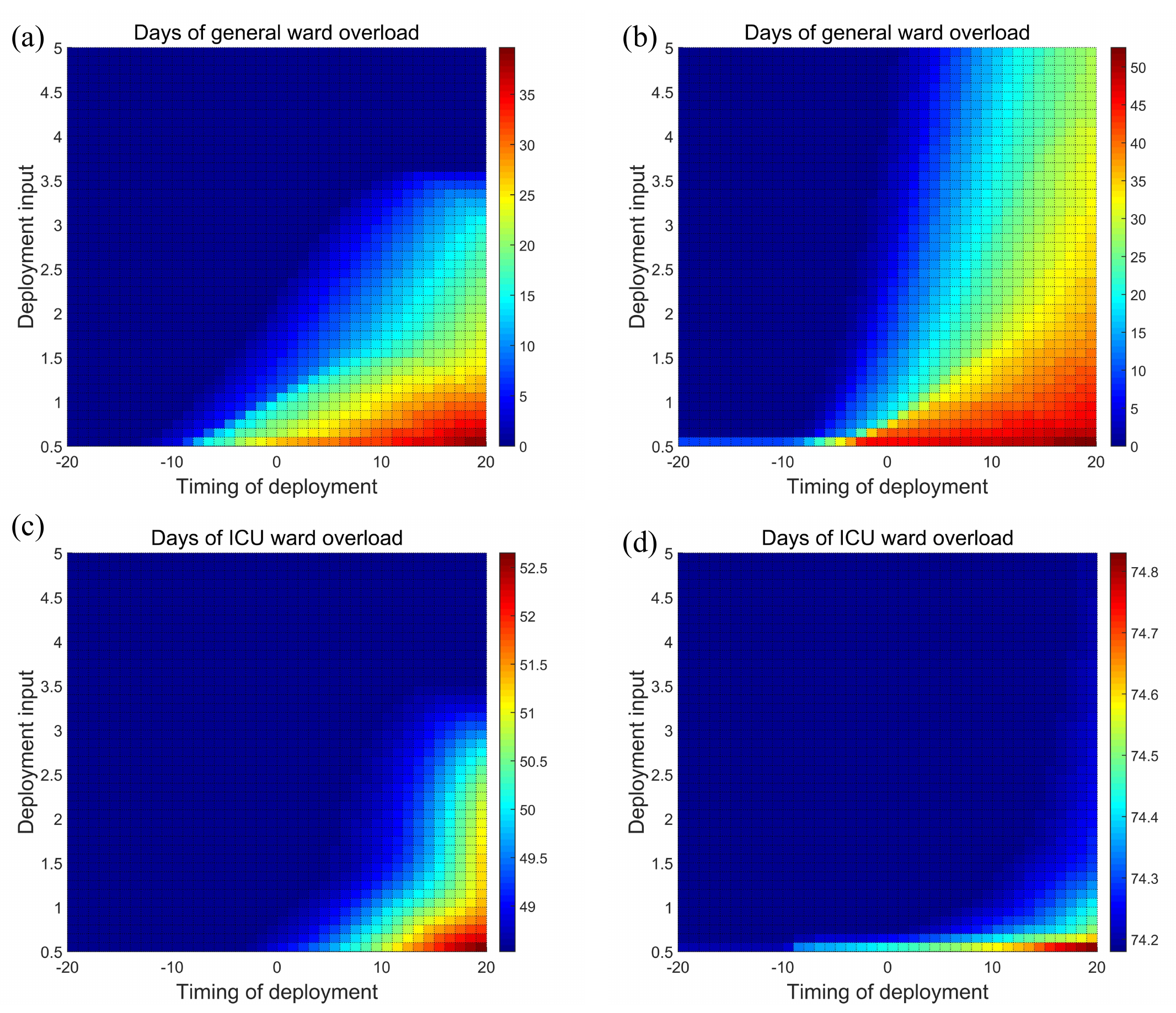}
\caption{ {\em Change in overload days for fixed ICU resource deployment
and varying GW resource deployment.} (a,b) Overload days of GW in Wuhan
and Lombardy, respectively. (c,d) Overload days of ICU in Wuhan and Lombardy,
respectively.}
\label{fig:S5}
\end{figure}

\begin{figure} [ht!]
\centering
\includegraphics[width=0.8\linewidth]{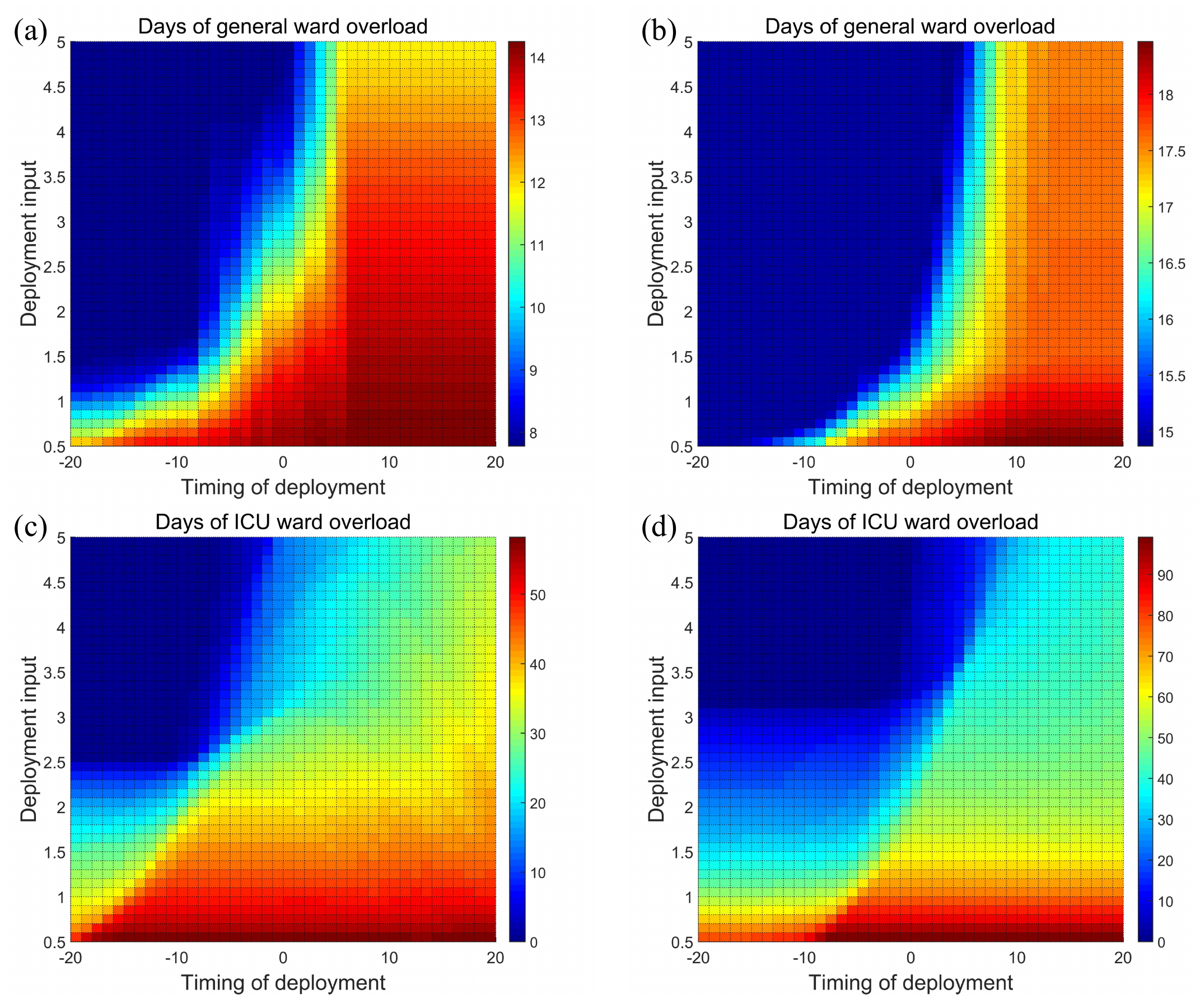}
\caption{ {\em Change in overload days for fixed GW resource deployment
and varying ICU resource deployment.} (a,b) Overload days of GW in Wuhan
and Lombardy, respectively. (c,d) Overload days of ICU in Wuhan and Lombardy,
respectively.}
\label{fig:S6}
\end{figure}

We study the impacts of varying both GW and ICU resources on the number of
deaths. As shown in Tab~\ref{tab:S7} and Fig.~\ref{fig:S4}, if resources
had been deployed one week in advance, the death toll in Wuhan and Lombardy
would have been reduced by 14\% and 3.4\%, respectively. If the input of
resources had been doubled, the death toll would have been suppressed by 22\%
and 15\%, respectively. If the deployment had been one week ahead of the
actual time and the input of resources had been doubled, the death toll
would have been reduced by 30\% and 17\% in Wuhan and Lombardy, respectively,
indicating that varying the two factors simultaneously can be more effective
at reducing the death toll. Figure~\ref{fig:S5} demonstrates the change in
the ward overload days when GW deployment is modified while ICU deployment
is unchanged. Figure~\ref{fig:S6} displays the change in the ward overload
days when GW deployment is unchanged but the ICU deployment plan is modified.

\newpage

\section*{\large Supplementary Note 4: Estimation of state transition fractions for different age groups}

In the main text, we divide the patients in Wuhan and Lombardy into three
age groups: [0-69], [70-79] and [80+], at intervals of 10\%. The transition
fractions from M state to F state, from G state to W state, from F state
to C state and the ICU mortality rates are different for different age groups.
We articulate a linear regression method to estimate the state transition
fractions for different age groups. Take as an example the transition
from  M state to F state. We first introduce the matrix of average transition
fraction, as (detailed in SN2)
\begin{equation}
\begin{aligned}
|P|_{M\rightarrow F}=\left(
  \begin{array}{ccc}
    |p|_1^{M\rightarrow F} \\
    \vdots \\
    |p|_1^{M\rightarrow F}\\
  \end{array}
\right),
\end{aligned}
\end{equation}
where $|p|_{i}^{M\rightarrow F}$ represents the average transition fraction
from M state to F state at stage $i$. The age distribution of patients in the
M state at different stages is
\begin{equation}
\begin{aligned}
D_{M}=\left(
  \begin{array}{ccc}
    d_{1,1}^{M}&\cdots &d_{1,y}^{M}  \\
    \vdots &\ddots &\vdots \\
    d_{s,1}^{M}&\cdots &d_{s,y}^{M}\\
  \end{array}
\right),
\end{aligned}
\end{equation}
where $d_{i,j}^{M}$ is the fraction of patients in age group $j$ among all
patients at stage $i$. We then define the matrix of average transition
fraction from M state to F state in different age groups as
%\begin{displaymath}
\begin{equation}
\begin{aligned}
R_{M\rightarrow F}=[r_1^{M\rightarrow F}\cdots r_y^{M\rightarrow F}],
\end{aligned}
\end{equation}
%\end{displaymath}
where $r_{j}^{M\rightarrow F}$ is the average transition fraction from
M state to F state in the $j$th age group during the whole recovery period.

\begin{figure} [ht!]
\centering
\includegraphics[width=\linewidth]{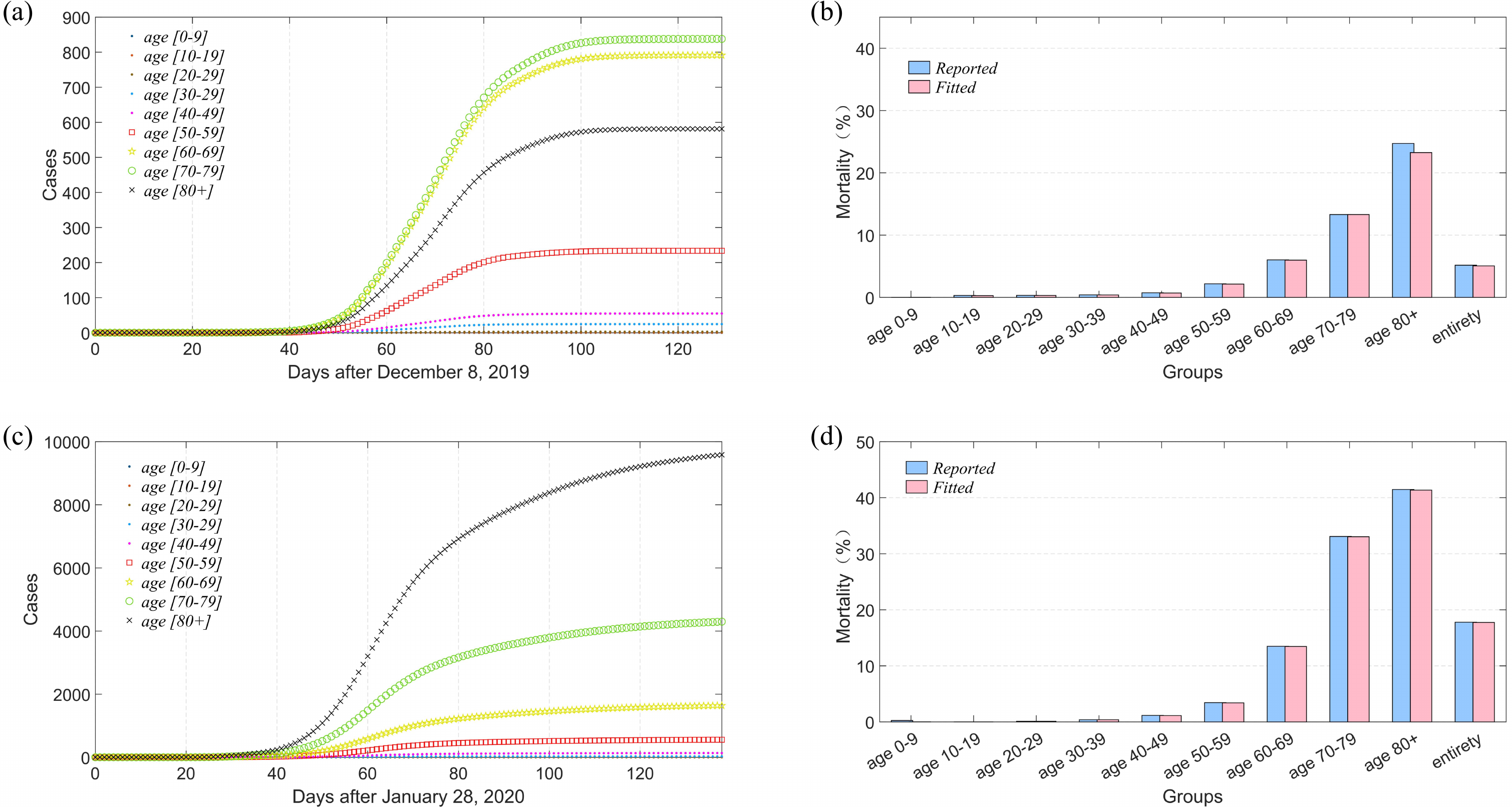}
\caption{ {\em Mortality rate of different age groups and evolution of the
number of deaths.} (a) Actual data of mortality rate of different age groups
in Wuhan versus model prediction. (b) Change in the number of deaths by age
groups in Wuhan. (c) Actual mortality data for all age groups in Lombardy
versus model prediction. (d) Number of deaths in Lombardy in each age group.
In (a) and (c), the legends from top to bottom are age [0-9] to [80].
In (b) and (d), the light blue and pink cylinders, respectively, represent
the actual and simulated mortalities in each group, where the $x$-axis marks
the groups with growing age.}
\label{fig:S7}
\end{figure}

The parameters can be obtained empirically. We make a linear transformation
of matrix $R_{M\rightarrow F}$ to obtain the estimation of the fraction of the
M-state patients to F-state whose weighted average transition fraction is
$|P|_{M\rightarrow F}$. The linear transformation $\mbox{LT}$ of different
stages is given by
\begin{equation}
\begin{aligned}
\mbox{LT}=\frac{|P|_{M\rightarrow F}}{D_{M}\cdot R_{M\rightarrow F}^T},
\end{aligned}
\end{equation}
Finally, the transition fractions from M state
to F state for different age groups are
\begin{equation}
\begin{aligned}
P_{M\rightarrow F}=LT\cdot \mbox{ext}(1,y)\cdot R_{M\rightarrow F}=\left(
  \begin{array}{ccc}
    p_{1,1}^{M\rightarrow F}&\cdots &p_{1,y}^{M\rightarrow F}  \\
    \vdots &\ddots &\vdots \\
    p_{s,1}^{M\rightarrow F}&\cdots &p_{s,y}^{M\rightarrow F}\\
  \end{array}
\right),
\end{aligned}
\end{equation}
where the rows represent stages, the columns represent groups, and
$p_{i,j}^{M\rightarrow F}$ denotes the transition fraction from M-state
to F-state in the $j$th age group at stage $i$. $\mbox{ext}(n,m)$ is an auxiliary matrix of $n$ rows and $m$ columns
whose elements are all one. The patient age distribution matrix $D_{F}$
is
\begin{equation}
\begin{aligned}
D_{F}=\left(
  \begin{array}{ccc}
    p_{1,1}^{M\rightarrow F}\cdot d_{1,1}^{M}&\cdots &p_{1,y}^{M\rightarrow F} \cdot d_{1,y}^{M} \\
    \vdots &\ddots &\vdots \\
    p_{s,1}^{M\rightarrow F}\cdot d_{s,1}^{M}&\cdots &p_{s,y}^{M\rightarrow F}\cdot d_{s,y}^{M}\\
  \end{array}
\right).
\end{aligned}
\end{equation}
Where $R$ and $D_M$ and $\mid P \mid$ are known (see Supplementary Table S3 and Supplementary Table S8.). Similarly, we can obtain the estimation of the transition fractions among
other states for different age groups at different stages. To simplify the
model, we assume that the age distribution of M-state patients at each stage is
identical($d_{1,j}^M=d_{2,j}^M=\ldots=d_{s,j}^M,j=1,2,\ldots,y$), as shown in Table~\ref{tab:S8}. The reference transition fraction
matrix among the states is also given in Table~\ref{tab:S8}. Finally, as shown in Fig.~\ref{fig:S7}, we divide the patient
population into nine groups.

\newpage

\section*{\large Supplementary Note 5: A scheme to realize different admission
strategies -- translation weighting}

We have developed a translation weighting scheme to achieve different admission
strategies. We set the priority weight of the $i$th age group as $w_i$ and
denote $\tau_i$ as the state age of the $i$th age group. After weighting, the
new state age $\tau_i'$ becomes,
\begin{equation}
\begin{aligned}
\tau_i'=\tau_i+w_i.
\end{aligned}
\end{equation}
The state age of patients in the age group $i$ is increased by $w_i$ units
(days). The patients are then admitted according to the weighted state age
under FCFS.

\begin{figure} [ht!]
\centering
\includegraphics[width=0.8\linewidth]{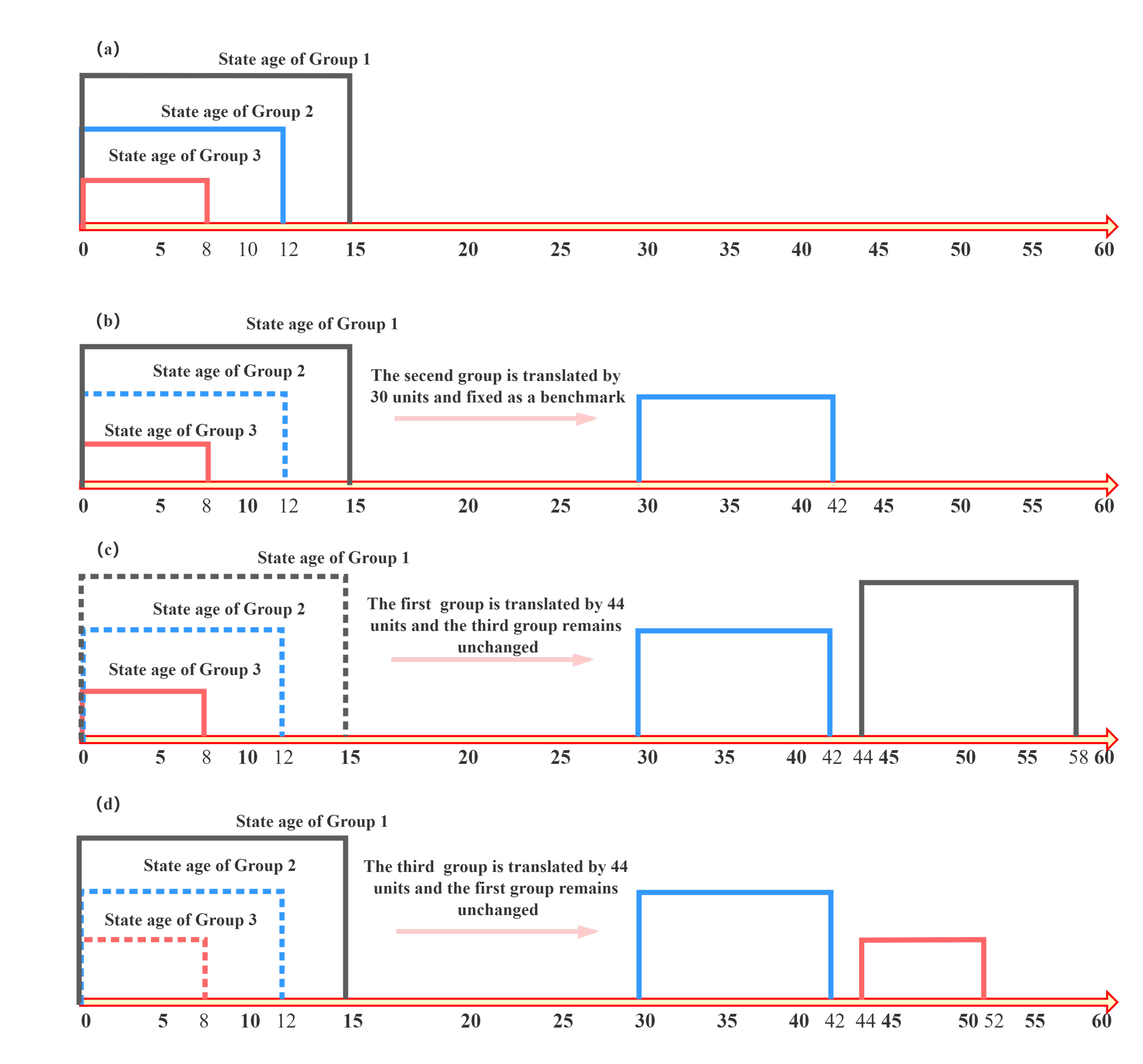}
\caption{ {\em Weighted state age of each age group.} (a) Initial state age
of the three groups of patients. (b) The state age of the second group is
translated by 30 units and fixed as a benchmark. (c) The state age of the
first group is translated by 44 units, and that of the third group remains
unchanged. (d) The state age of the third group is translated by 44 units,
and that of the first group remains unchanged.}
\label{fig:S8}
\end{figure}

We divide the patients into three age groups, which are patients younger
than 70 years old, patients between 70 and 80 years old, and patients older
than 80 years. Figure~\ref{fig:S8} presents examples of several weight
combinations. For example, in Fig.~\ref{fig:S8}(b), the weights of the three
age groups are $w_1=0$, $w_2=30$ and $w_3=0$, respectively. The state ages
of the three age groups after weighting are also illustrated.

\section*{\large Supplementary Note 6: Medical resources per capita and the age
distribution of confirmed cases in four countries}

We collect data on ward resources per 100,000 people per capita in China,
South Korea, Italy and Spain, as well as the fractions of COVID-19 confirmed
cases in each age group, as shown in Tables~\ref{tab:S9} and \ref{tab:S10}.

\clearpage

\section*{\large Supplementary Tables}

\begin{table}[!htbp]
\centering
\caption{State transition time delay parameter setting}
\resizebox{470pt}{!}{
\begin{tabular}{|c|c|c|c|c|}
\hline
\multicolumn{2}{|c|}{Parameter(delay)} &Estimate/assumption&Definition&Justification\\
\hline
$\mu_{M\rightarrow R}$ & Wuhan/Lombard&14d&Average delay from M state to report recovery&\cite{label1}\\
\hline
$\mu_{M\rightarrow F}$&Wuhan/Lombard&5d(std=6.67)& Average delay from M state to F state &\cite{label15}\\
\hline
$\mu_{F\rightarrow M_X}$&Wuhan/Lombard&8d(std=4.4477)& Average delay from F state to $M_X$ state&\cite{label10}\\
\hline
$\mu_{F\rightarrow C}$&Wuhan/Lombard&2d(std=3.7064)& Average delay from F state to C state &\cite{label15, label25}\\
\hline
$\mu_{G\rightarrow M_X}$&Wuhan/Lombard& 8d(std=4.4477) &Average delay from G state to $M_X$ state&\cite{label10}\\
\hline
$\mu_{G\rightarrow W}$&Wuhan/Lombard&3d(std=3.7064) &Average delay from G state to W state &\cite{label15}\\
\hline
$\mu_{U\rightarrow G_U}$&Wuhan/Lombard&8d(std=4.4477)& Average delay from U state to $G_U$ state &\cite{label16}\\
\hline
$\mu_{U\rightarrow D}$&Wuhan/Lombard&7d(std=5.93) &Average delay from U state to D state &\cite{label16, label17}\\
\hline
$\mu_{C\rightarrow D}$&Wuhan/Lombard &1d(std=2.2239)&Average delay from C state to D state &Assume\\
\hline
$\mu_{W\rightarrow D}$&Wuhan/Lombard&3d(std=3.7064) &Average delay from W state to D state &Assume\\
\hline
$\mu_{G_U\rightarrow M_X}$&Wuhan/Lombard&8d(std=4.4477) &Average delay from $G_U$ state to $M_X$ state &\cite{label10}\\
\hline
$\mu_{M_X\rightarrow R}$ & Wuhan/Lombard&14d&Average delay from $M_X$ state to report recovery&\cite{label1}\\
\hline
\end{tabular}}
\label{tab:S1}
\end{table}

\begin{table}[!htbp]
\centering
\caption{Parameter setting of state transition fraction}
\resizebox{470pt}{!}{
\begin{tabular}{|c|c|c|c|c|}
\hline
\multicolumn{2}{|c|}{Parameter(delay)} &Estimate/assumption&Definition&Justification\\
\cline{1-4}
$P_{M\rightarrow F}$&Wuhan/Lombard&see Tab S3.1 and S3.2 & Transition fraction from M state to F state &\\
\hline
\multirow{2}*{$P_{F\rightarrow C}$} & Wuhan&48.3\% (41.06\%, 54.97\%)&\multirow{2}*{Transition fraction from F state to C state}&\multirow{4}*{Fitting according to the reported death data}\\
\cline{2-3}
&Lombard&88.86\% (76.8\%, 100\%)&&\\
\cline{1-4}
\multirow{2}*{$P_{G\rightarrow W}$} & Wuhan&27.4\% (26.24\%, 28.51\%)&\multirow{2}*{Transition fraction from G state to W state}&\\
\cline{2-3}
&Lombard&63.77\% (62.9\%, 64.71\%)&&\\
\hline
$P_{U\rightarrow D}$&Wuhan/Lombard&61.5\% & Transition fraction from U state to D state &\cite{label16, label26}\\
\hline
\end{tabular}}
\label{tab:S2}
\end{table}

\begin{table}[!htbp]
\centering
\caption{Setting of transition fraction from M state to F state in five stages in Wuhan}
\resizebox{470pt}{!}{
\begin{tabular}{|c|c|c|c|c|c|}
\hline
Date& Dec 8, 2019-Jan 9, 2020& Jan 10-Jan 22& Jan 23-Feb 1& Feb 2-Feb 16& Feb 17--\\
\hline
$P_{M\rightarrow F}(Wuhan)$&53.10\%& 35.10\%& 23.50\%& 15.90\%& 10.30\%\\
\hline
Justification&\multicolumn{5}{|c|}{\cite{label27}}\\
\hline
\end{tabular}}
\label{tab:S3}
\end{table}

\begin{table}[!htbp]
\centering
\caption{Setting of transition fraction from M state to F state in four stages in Lombardy}
\resizebox{470pt}{!}{
\begin{tabular}{|c|c|c|c|c|}
\hline
Date& Jan 28, 2020-Feb 19, 2020& Feb 20-Feb 25& Feb 26-Mar 20 1& Mar 21--\\
\hline
$P_{M\rightarrow F}(Lombardy)$&63.00\%& 61.00\%& 56.00\%& 25\%(24.29\%, 25.65\%)\\
\hline
Justification&\multicolumn{3}{|c|}{\cite{label28}}&Fitting according  to the reported death data\\
\hline
\end{tabular}}
\label{tab:S4}
\end{table}

\begin{table}[!htbp]
\centering
\caption{Actual medical resource deployment plan for Wuhan, China~\cite{label30}}
\resizebox{200pt}{!}{
\begin{tabular}{|c|c|}
\hline
Date& Beds(ICU+General ward)\\
\hline
2020/2/1&842\\
\hline
2020/2/2&842\\
\hline
2020/2/3&1962\\
\hline
2020/2/4&2234\\
\hline
2020/2/5&2264\\
\hline
2020/2/6&2515\\
\hline
2020/2/7&2657\\
\hline
2020/2/8&2911\\
\hline
2020/2/9&3263\\
\hline
2020/2/10&4307\\
\hline
2020/2/11&4461\\
\hline
2020/2/12&5873\\
\hline
2020/2/13&6078\\
\hline
2020/2/14&6636\\
\hline
2020/2/15&6926\\
\hline
2020/2/16&7035\\
\hline
2020/2/17&7067\\
\hline
2020/2/18&7225\\
\hline
2020/2/19&7296\\
\hline
2020/2/20&7560\\
\hline
2020/2/21&7560\\
\hline
2020/2/22&7844\\
\hline
2020/2/23&7844\\
\hline
2020/2/24&7936\\
\hline
2020/2/25&8194\\
\hline
\end{tabular}}
\label{tab:S5}
\end{table}

\begin{table}[!htbp]
\centering
\caption{Actual medical resource deployment in Lombardy, Italy~\cite{label34}}
\resizebox{200pt}{!}{
\begin{tabular}{|c|c|c|}
\hline
Date& Beds(ICU)& Beds(General ward)\\
\hline
2020/2/24&19&76\\
\hline
2020/2/25&25&79\\
\hline
2020/2/26&25&79\\
\hline
2020/2/27&41&172\\
\hline
2020/2/28&47&235\\
\hline
2020/2/29&80&256\\
\hline
2020/3/1&106&406\\
\hline
2020/3/2&127&478\\
\hline
2020/3/3&167&698\\
\hline
2020/3/4&209&877\\
\hline
2020/3/5&244&1169\\
\hline
2020/3/6&309&1622\\
\hline
2020/3/7&359&1661\\
\hline
2020/3/8&399&2217\\
\hline
2020/3/9&440&2802\\
\hline
2020/3/10&466&3319\\
\hline
2020/3/11&560&3852\\
\hline
2020/3/12&605&4247\\
\hline
2020/3/13&650&4435\\
\hline
2020/3/14&732&4898\\
\hline
2020/3/15&767&5500\\
\hline
2020/3/16&823&6171\\
\hline
2020/3/17&879&6953\\
\hline
2020/3/18&924&7285\\
\hline
2020/3/19&1006&7387\\
\hline
2020/3/20&1050&7735\\
\hline
2020/3/21&1093&8258\\
\hline
2020/3/22&1142&9439\\
\hline
2020/3/23&1183&9266\\
\hline
2020/3/24&1194&9711\\
\hline
2020/3/25&1236&10026\\
\hline
2020/3/26&1263&10681\\
\hline
2020/3/27&1292&11137\\
\hline
2020/3/28&1319&11152\\
\hline
2020/3/29&1328&11613\\
\hline
2020/3/30&1330&11815\\
\hline
2020/3/31&1334&11883\\
\hline
2020/4/1&1342&12009\\
\hline
2020/4/2&1351&12009\\
\hline
2020/4/3&1381&12009\\
\hline
\end{tabular}}
\label{tab:S6}
\end{table}

\begin{table}[!htbp]
\centering
\caption{Comparison of different medical resource deployment plans
(simultaneously adjusting GW and ICU resources)}
\resizebox{400pt}{!}{
\begin{tabular}{|c|c|c|}
\hline
\multirow{2}*{}&\multicolumn{2}{|c|}{Simulated death toll}\\
\cline{2-3}
&Wuhan&Lombard\\
\hline
The actual deployment& 2553& 16363\\
\hline
Sufficient resources&Decrease by 33\%&Decrease by 22\%\\
\hline
Deploy 7 days in advance&Decrease by 14\%&Decrease by 3.4\%\\
\hline
Deploy 14 days in advance&Decrease by 18\%&Decrease by 4\%\\
\hline
7 days delay in deployment&Increase by 36\%&Increase by 14\%\\
\hline
14 days delay in deployment&Increase by 57\%&Increase by 28\%\\
\hline
Resource input *0.5&Increase by 43\%&Increase by 26\%\\
\hline
Resource input *2&Decrease by 22\%&Decrease by 15\%\\
\hline
Resource input *5&Decrease by 32\%&Decrease by 22\%\\
\hline
Deploy 7 days in advance and invest resources *2&Decrease by 30\%&Decrease by 17\%\\
\hline
\end{tabular}}
\label{tab:S7}
\end{table}

\begin{table}[!htbp]
\centering
\caption{Reference transition fraction matrix among states and age distribution of patients in M state}
\resizebox{470pt}{!}{
\begin{tabular}{|c|c|c|c|c|}
\hline
\multicolumn{2}{|c|}{Parameter(Reference matrix)} &Estimate/assumption&Definition&Justification\\
\hline
\multirow{2}*{$R_{M\rightarrow F}$} & Wuhan&[0.0953 0.2430 0.2730]&\multirow{2}*{Reference transformation matrix from M state to F state}&\multirow{2}*{\cite{label10}}\\
\cline{2-3}
&Lombard&[0.0827 0.2430 0.2730]&&\\
\hline
\multirow{2}*{$R_{F\rightarrow C/G\rightarrow W}$} & Wuhan&[0.0982 0.1419 0.2393]&\multirow{2}*{Reference transformation matrix from F/G state to C/W state}&\multirow{2}*{Fitting according to the reported death data fitting}\\
\cline{2-3}
&Lombard&[0.1249 0.2240 0.2393]&&\\
\hline
$R_{U\rightarrow D}$&Wuhan/Lombard&[0.2297 0.7727 0.8571] &Reference transformation matrix from U state to D state &\cite{label17}\\
\hline
\multirow{2}*{$D_{M}$} & Wuhan&[0.8240 0.1260 0.0500]&\multirow{2}*{Age group distribution matrix of M-state patients}&\multirow{2}*{\cite{label19}}\\
\cline{2-3}
&Lombard&[0.6050 0.1420 0.2530]&&\\
\hline
\end{tabular}}
\label{tab:S8}
\end{table}

\begin{table}[!htbp]
\centering
\caption{ Average wards per 100,000 people}
\resizebox{470pt}{!}{
\begin{tabular}{|c|c|c|}
\hline
Country/Resource&General ward beds per 100000 people&ICU beds per 100000  people\\
\hline
China&400&3.6\\
\hline
Korea&530&10.6\\
\hline
Italy&333&12.5\\
\hline
Spain&269&9.7\\
\hline
Justification&\multicolumn{2}{|c|}{\cite{label38}}\\
\hline
\end{tabular}}
\label{tab:S9}
\end{table}

\begin{table}[!htbp]
\centering
\caption{ Proportion of confirmed cases in each age group in each country}
\resizebox{470pt}{!}{
\begin{tabular}{|c|c|c|c|c|}
\hline
\multicolumn{5}{|c|}{Age distribution of confirmed cases}\\
\hline
Groups/Country&China&Italy&Korea&Spain\\
\hline
0-9&0.009312&0.006245&0.012344&0.002639\\
\hline
10-19&0.01229&0.008122&0.052919&0.005433\\
\hline
20-29&0.081013&0.04061&0.272823&0.05018\\
\hline
30-39&0.170129&0.069164&0.106507&0.096366\\
\hline
40-49&0.191865&0.128128&0.133589&0.150725\\
\hline
50-59&0.224033&0.198045&0.183732&0.186374\\
\hline
60-69&0.192134&0.173838&0.126316&0.165787\\
\hline
70-79&0.087706&0.185173&0.066411&0.158476\\
\hline
80+&0.031519&0.188301&0.045359&0.18402\\
\hline
Justification&\multicolumn{4}{|c|}{\cite{label39}}\\
\hline
\end{tabular}}
\label{tab:S10}
\end{table}

\clearpage

\section*{\large Supplementary Figures}

\clearpage

\bibliographystyle{naturemag}
\bibliography{COVID_19}

\end{document}